\documentclass[12pt]{article}

\usepackage[english]{babel}

\usepackage[letterpaper,top=1in,bottom=1in,left=1in,right=1in,marginparwidth=1.75cm]{geometry}

\usepackage[T1]{fontenc}
\usepackage{lmodern}
\usepackage{setspace}
\usepackage{amsmath,amssymb,amsfonts}
\usepackage{mathtools}
\usepackage{bm,bbm}
\usepackage{mathrsfs}

\usepackage{graphicx,epstopdf}
\usepackage{caption,subcaption}
\usepackage{tikz}
\usetikzlibrary{bayesnet}
\usepackage{float}
\usepackage{rotating}

\usepackage{xcolor}
\usepackage[colorlinks=true,allcolors=blue]{hyperref}

\usepackage[inline]{enumitem}      
\usepackage{enumerate}             

\newlist{todolist}{itemize}{2}
\setlist[todolist]{label=$\square$}

\usepackage{authblk}               
\usepackage[normalem]{ulem}        
\usepackage{verbatim}
\usepackage{standalone}
\usepackage{booktabs,dcolumn}      
\usepackage{tabularx}
\usepackage{longtable}
\usepackage{bibunits}
\usepackage[ruled,vlined]{algorithm2e}  
\usepackage{derivative}
\usepackage{microtype}
\usepackage{indentfirst}
\usepackage{lipsum}                
\usepackage{chngcntr}
\usepackage{cleveref}
\usepackage{multirow}

\usepackage{soul}

\usepackage[toc,page]{appendix}
\usepackage{footnote}
\makesavenoteenv{tabular}
\makesavenoteenv{table}

\usepackage{natbib}
\bibpunct[, ]{(}{)}{,}{a}{}{,}%
\newlength{\bibitemsep}
\newlength{\bibparskip}
\let\oldthebibliography\thebibliography
\renewcommand\thebibliography[1]{%
  \oldthebibliography{#1}%
  \setlength{\parskip}{-0.5\bibitemsep}%
  \setlength{\itemsep}{-0.5\bibparskip}%
}

\newcolumntype{.}{D{.}{.}{-1}}
\newcolumntype{+}{>{\global\let\currentrowstyle\relax}}
\newcolumntype{^}{>{\currentrowstyle}}

\usepackage{amsthm}

\theoremstyle{remark}

\mathchardef\mhyphen="2D

\title{A Bayesian Multi-State Data Integration Approach for Estimating County-level Prevalence of Opioid Misuse in the United States}
\author[1]{Zixuan Feng}
\author[1]{Qiushi Chen\thanks{Corresponding author. Email: q.chen@psu.edu}}
\author[1]{Paul Griffin}
\affil[1]{Harold and Inge Marcus Department of Industrial and Manufacturing Engineering, The Pennsylvania State University}
\author[2]{Le Bao}
\affil[2]{Department of Statistics, The Pennsylvania State University}
 
\date{}

\begin{document}
\maketitle

\begin{abstract}
Drug overdose deaths, including from opioids, remain a significant public health threat to the United States (US). To abate the harms of opioid misuse, understanding its prevalence at the local level is crucial for stakeholders in communities to develop response strategies that effectively use limited resources. Although there exist several state-specific studies that provide county-level prevalence estimates, such estimates are not widely available across the country, as the datasets used in these studies are not always readily available in other states, which, therefore, has limited the wider applications of existing models. To fill this gap, we propose a Bayesian multi-state data integration approach that fully utilizes publicly available data sources to estimate county-level opioid misuse prevalence for all counties in the US. The hierarchical structure jointly models opioid misuse prevalence and overdose death outcomes, leverages existing county-level prevalence estimates in limited states and state-level estimates from national surveys, and accounts for heterogeneity across counties and states with counties’ covariates and mixed effects. Furthermore, our parsimonious and generalizable modeling framework employs horseshoe+ prior to flexibly shrink coefficients and prevent overfitting, ensuring adaptability as new county-level prevalence data in additional states become available. Using real-world data, our model shows high estimation accuracy through cross-validation and provides nationwide county-level estimates of opioid misuse for the first time.
\end{abstract}


\section{Introduction}
The opioid epidemic in the United States (US) has become one of the most significant public health threats in the past two decades \citep{Salmond2019-qq, hebert2024impact}. It has resulted in over 1 million drug overdose deaths since 1999 \citep{KennedyHendricks2024} and has been evolving through multiple waves, starting with prescription opioids, followed by heroin, synthetic opioids \citep{ciccarone2019triple}, and now polysubstance use involving opioids and stimulants, which is referred to as the ``fourth wave'' \citep{Ciccarone2021-uc}. In addition to the devastating mortality burden, the opioid epidemic has had a substantial economic burden on healthcare, criminal justice, and lost productivity \citep{Luo2021-gy, florence2021economic}, with the total economic toll of the opioid epidemic estimated to reach nearly \$1.5 trillion in 2020 alone \citep{opioidcost2022}. Despite a recent decline in overdose deaths, the total mortality burden of drug overdoses remains staggering: over 80,000 people died from drug overdoses in the year 2024 alone, with approximately 70\% involving opioids \citep{cdc2025press}. 

To abate the harms of opioid and substance misuse, understanding the needs and burden at the community level is crucial for local stakeholders to develop effective response strategies utilizing limited resources through various interventions  \citep{ElBassel2021, SAMHSA2022CommunityEngagement}. One important measure is the {\em prevalence} of people who misuse opioids and are at risk of opioid and drug overdose deaths. The common source for nationwide substance use-related prevalence estimates is the National Survey on Drug Use and Health (NSDUH), one of the largest and most comprehensive annual general population surveys in the US focusing on substance use and mental health \citep{nsduh2018}. While NSDUH routinely publishes estimates for the prevalence of opioid and drug use at the national, state, and sub-state regional levels \citep{samhsa2023national}, estimates at the county level are not generally available (except for a few large counties considered as standalone sub-state regions), since limited samples for relatively rare events (e.g., substance use) in each county are collected in the multi-stage sampling design of the national survey \citep{samhsa2024county}. As national opioid settlement funds have started to be disbursed to counties across the nation to help abate the harms of the local opioid crisis \citep{RHUBART2022, Monnat2018,Skinner2024}, estimates of county-level prevalence of opioid misuse are needed more than ever to help local policymakers understand unmet needs in communities. 

Due to the lack of nationwide county-level prevalence estimates for opioid misuse, several recent studies have developed state-specific models for county-level estimation that required state-specific datasets. For example, \citet{Hepler2023-mu} generate county-level estimates of opioid misuse in Ohio during 2007-2019 using a Bayesian hierarchical spatiotemporal model that leverages the number of overdose deaths and treatment admissions in each county of Ohio to redistribute the state-level prevalence estimates from NSDUH. \citet{SantaellaTenorio2024} extend this model by incorporating opioid overdose emergency department (ED) visits, in addition to overdose deaths and treatment data, to estimate the opioid misuse prevalence for counties of New York State during 2007-2018. \citet{kline2025estimating} apply a similar abundance model to North Carolina counties based on county-level observations of illicit opioid overdose deaths and treatment data. In addition, \citet{Doogan2022-lf} estimate the opioid use disorder (OUD) prevalence among the Medicaid population in 19 Ohio counties based on individual-level Medicaid administrative claims and Vital Statistics data. The capture-recapture method has also been applied to estimate county-level OUD prevalence in Massachusetts \citep{Barocas2018-ke} and Kentucky \citep{Thompson2023}. 

However, the methods of estimating opioid misuse prevalence in existing studies are not universally applicable to all states. The state-specific statistical models rely on {\em ad-hoc} measures collected from sources or programs (such as ED visits, claims data, or substance use treatment programs) that are specific to the respective state, making it challenging to replicate the model in other states. The capture-recapture analysis requires individually linked databases; it is feasible only if the state has already built an extensive data infrastructure to support it, which, however, is not typically available. The multiplier method, although requiring less of the individually linked datasets, is often used to adjust for prevalence estimates at a more aggregate level (e.g., at the state or national level \citep{Keyes2022-zv, Mojtabai2022}) and has a limited capability to capture the intricate heterogeneity at a more granular level.

To fill the gap in the literature, we propose a method of estimating the prevalence of opioid misuse at the county level that is not restricted to state-specific data sources and is applicable nationwide. We develop a Bayesian multi-state data integration approach with two key data principles in mind. First, we incorporate the best available evidence on the prevalence of opioid misuse at multiple levels, including county-level estimates that are based on epidemiological studies (e.g., capture-recapture study) in certain states, and state-level estimates from NSDUH for all states, while recognizing the NSDUH estimates are subject to potential underestimation \citep{Reuter2021-ly}. Second, our approach fully utilizes publicly available data sources that are related to opioid misuse and readily available at the county level for all states, including the Centers for Disease Control and Prevention (CDC) Wide-ranging ONline Data for Epidemiologic Research (WONDER) for drug overdose deaths \citep{cdc2024wonder}, US Census data \citep{uscensus} and County Health Rankings \citep{countyhealth2024} for demographic and socioeconomic information at the county-level. Our proposed method shares a common conceptualization of the estimation problem with the existing state-specific integrated abundance models, such as \citet{Hepler2023-mu} and \citet{kline2025estimating}, that leads to the integration of multiple data sources through Bayesian hierarchical models. The key distinctions of our approach are the integration of {\em prevalence data} at different geographical levels (i.e., state and county) from multiple sources and the reliance only on publicly available data, enabling us to infer county-level prevalence across the country beyond only limited states. 

We believe that this is the first model to jointly estimate county-level prevalence of opioid misuse in all US states, which fills a critical knowledge gap in the substance use-related domain.  The main contributions of our Bayesian multi-state data integration approach are summarized as follows.

\begin{itemize}
    \item {\em Data integration}: It seamlessly integrates a diverse set of publicly available data sources, including state-level prevalence data from NSDUH, county-level prevalence estimates from published state-specific studies, overdose mortality, and county-level demographic and socioeconomic characteristics, to infer the opioid misuse prevalence at the county level. Our model is tailored to capture underestimation in existing data sources and accommodate data suppression common in public datasets. We also explicitly formulate the link between opioid misuse and OUD to reconcile the definitional differences, making our approach flexible for integrating prevalence estimates reported differently from various data sources. 
    \item {\em Hierarchical structure}: It simultaneously models opioid misuse prevalence and opioid overdose death outcomes, and adjusts for the NSDUH underestimation for all states. It explicitly models the heterogeneity of opioid misuse prevalence across counties and states with a county's covariates and mixed effects. 
    \item {\em Parsimonious and generalizable model framework}: It provides a generalizable modeling framework for systematic estimation of opioid misuse prevalence that can easily incorporate additional covariate data. It applies horseshoe+ prior, a Bayesian prior that allows for the flexible shrinkage of coefficients, to prevent overfitting. This model can also be naturally extended as new evidence emerges, such as county-level prevalence estimates for additional states, which may become available from new published studies in the future.     
\end{itemize}

The remainder of this paper is organized as follows. Section \ref{sec:data} describes the opioid and substance use-related datasets that are widely available, which motivate and guide the structure of our proposed statistical model. Section \ref{sec:model} presents our Bayesian hierarchical model for estimating county-level opioid misuse prevalence in all US states and describes our computational method. Section \ref{sec:results} shows model diagnostics and validation using real-world data, and the results of prevalence estimation for all 50 states. Section \ref{sec: discussion} concludes with discussions of key findings, practical implications, and future directions.

\section{Data}
\label{sec:data}
In this section, we describe the data used in our model development. They are commonly used data sources in opioid and substance use research and are also publicly available for all states, which motivates the development of our Bayesian hierarchical model. 

\textbf{County population:} This serves as the base population for opioid misuse, which further leads to opioid overdose deaths. We use 5-year estimates of total population for each county from the American Community Survey for maximum coverage, as single-year estimates of county population are only available for those with populations over 65,000 \citep{uscensus}.

\textbf{Opioid-related overdose deaths:} The annual number of opioid-related overdose deaths for each county can be queried from the CDC WONDER Multiple Cause of Deaths database \citep{cdc2024wonder}, an extensively used data source in opioid and substance use-related health policy research \citep{Nam2017-ey, Buonora2022, FeuersteinSimon2020, McClellan2018}. Though the data are publicly accessible through online queries, any value below 10 in the query results is suppressed to protect individual confidentiality. Our preliminary analysis finds that about 45-70\% of the counties across the country have suppressed values over 2010-2021. Since the data suppression issue is not negligible, it will be explicitly accounted for in our model formulation. 

\textbf{Existing prevalence estimates:} County-level prevalence estimates for OUD have been reported in published capture-recapture studies for Massachusetts in 2011-2015 \citep{Barocas2018-ke} and Kentucky in 2017-2018 \citep{Thompson2023}. These county-level estimates can serve as reference points, allowing us to further model their relations with other county-level and state-level variables. We also include the NSDUH state-level estimates to capture relative differences in prevalence across the states and the link between opioid misuse and OUD. We acknowledge the underestimation in NSDUH estimates when using the data \citep{Reuter2021-ly,Keyes2022-zv} and thus systematically adjust for this data limitation in our model. 

\textbf{Opioid dispensing rate:} The number of opioid prescriptions dispensed per 100 population in each county in each year \citep{CDCDispensingRates} captures the availability of prescription opioids. Overprescribing of opioids helped fuel the opioid epidemic in its early years, and previous research has shown that higher dispensing rates were linked to increased misuse and overdose risks \citep{Grigoras2017, Sawyer2021}. 

\textbf{County socio-demographic factors:} The opioid epidemic should be understood within a social-ecological framework that emphasizes the interplay of individual, community, and societal influences on health outcomes \citep{Jalali2020OpioidCrisisFramework}. Previous studies have examined a wide range of demographic, social, economic, and health-related factors as county-level predictors for drug overdose mortality \citep{Cano2023-cf} and have used them in prevalence estimation \citep{Hepler2023-mu, SantaellaTenorio2024, Doogan2022-lf}. We obtain county socio-demographic factors from the County Health Rankings, a comprehensive dataset that consists of more than 90 measures across health behaviors, clinical care, social and economic factors, and physical environment domains \citep{countyhealth2024}. We select measures for education, unemployment, risk behavior, and access to health services that are broadly relevant to substance use from this dataset as county-level covariates. More detailed descriptions of the included covariates are provided later in Section \ref{sec:results} for the case study. 

\section{Methods}
\label{sec:model}

In this section, we present the multi-state Bayesian hierarchical model (BHM) for estimating county-level prevalence of opioid misuse using data from multiple data sources. We describe the model specification and computational approach in detail below. 

\subsection{Bayesian hierarchical model}
\label{sec:bhm}

\begin{figure}[h!]
\centering
\includegraphics[width=0.95\linewidth]{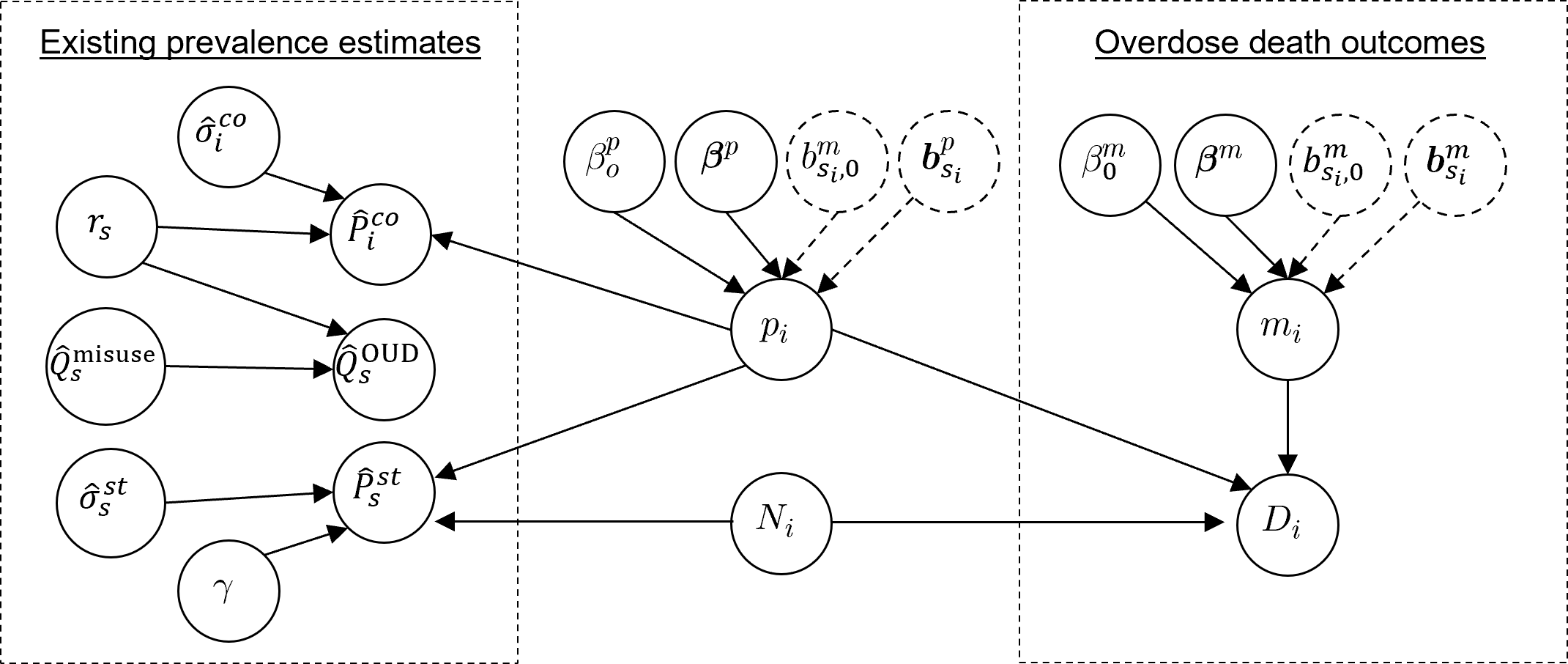}
\caption{Schematic of the proposed Bayesian hierarchical model. Observed quantities are in capital letters, and model parameters are in lower-case letters. The inclusion of possible random effects, as illustrated in dashed circles, is guided by residual analyses. Covariates and model priors are omitted for simplicity of the presentation.}
\label{fig:generative-graph}
\end{figure}

Our Bayesian hierarchical model consists of two parts: (1) models for directly observed numbers of opioid overdose deaths from the population with opioid misuse, and (2) models for existing available prevalence estimates related to opioid use at the county level (for some states) and state level (for all states). The overall model structure is illustrated in Figure \ref{fig:generative-graph}.

\subsubsection{Model overdose deaths}
Consider a set of counties $\mathcal{I}$ with total population $N_i$ for $i\in\mathcal{I}$. Let $p_i$ represent the proportion of county $i$'s population with opioid misuse (i.e., prevalence rate), and $m_i$ represent the probability of opioid overdose deaths (i.e., mortality rate) among the opioid misuse population in county $i$. The prevalence and mortality rates, both measured as percentages, are parameterized using the inverse-logit transformation as follows:
\begin{align}
    \text{logit}(p_i) = & \beta_0^p+ \bm{X}_i^{p}\bm{\beta}^{p},
    \label{eq:p_oud} \\
    \text{logit}(m_i) = & \beta_0^m + \bm{X}_i^{m}\bm{\beta}^{m}.
    \label{eq:q_death}
\end{align}
where $\bm{X}_i^p=(X_{i,1}^p, X_{i,2}^p, \cdots, X_{i,P}^p)$ are the set of covariates for county $i$, referred to as the {\em prevalence determinants}, $\beta_0^p$ and $\bm{\beta}^p$ are the fixed effect intercept and slope for the prevalence determinants $\bm{X}_i^p$, respectively. Similarly, $\bm{X}_i^m=(X_{i,1}^m, X_{i,2}^m, \cdots, X_{i, M}^m)$, $\beta_0^m$, and $\bm{\beta}^m$ represent the {\em mortality determinants}, fixed effect intercept, and slope for the mortality determinants, respectively.

We model the number of overdose deaths as a two-stage process. We first define an auxiliary variable $O_i$ to denote the number of individuals with opioid misuse in county $i$, where $O_i \mid N_i, p_i \sim \text{Poisson}(p_i N_i)$. Given the mortality rate $m_i$ in the opioid misuse population $O_i$, it follows that the number of overdose deaths $D_i \mid O_i, m_i \sim \text{Binomial}(O_i, m_i)$. The Poisson-Binomial models are equivalent to splitting a Poisson process with a splitting probability $m_i$, resulting in a Poisson distribution for overdose deaths $D_i$, which has an expected value of $m_i \cdot (p_i N_i)$. Therefore, we model the number of opioid overdose deaths $D_i$ with a Poisson distribution:
\begin{equation}
\label{eq:model-mortality}
    D_i \mid N_i, m_i, p_i\sim \text{Poisson}(m_i p_i N_i),
\end{equation}
which are conditionally independent across counties. Suppressed observations of overdose deaths are treated as left-censored and contribute $P(D_i\le c)$ under the Poisson model.  


\subsubsection{Incorporate existing prevalence estimates} 

There are prevalence estimates related to opioid use and misuse available from publicly accessible sources such as published literature and reports. Although these estimates are either limited to certain counties (e.g., from a state-specific capture-recapture study) or only available at an aggregate level (e.g., state-level reports by NSDUH), they may still contain valuable information that can be used to inform the prevalence estimation for all counties. It is worth noting that definitions of the prevalence may differ among these existing estimates, with some focusing on opioid misuse while others are restricted to OUD as a specific clinical condition that meets certain diagnostic criteria. In the following, we incorporate multiple sets of existing prevalence estimates into our model, including the county-level estimates for limited counties and the state-level estimates for all states.

{\bf County-level estimates:} Let $\mathcal{I}_e \subset \mathcal{I}$ denote the counties with existing county-level prevalence estimates for the OUD population from published epidemiological studies. Because opioid misuse is the focus of our study and encompasses a broader construct than OUD, which is defined by more restrictive clinical criteria, we model the population with OUD as a subset of those with opioid misuse. In particular, we define a parameter $r_s$ to represent the proportion of the population with opioid misuse meeting the criteria of OUD in state $s\in\mathcal{S}$. These OUD-to-opioid-misuse ratios $\{r_s\}_{s\in\mathcal{S}}$ are state-specific because the data from NSDUH reports to infer these parameters are only available at the state level. Let $\hat{Q}_s^{\text{OUD}}$ and $\hat{Q}_s^{\text{misuse}}$ denote the state-level counts of individuals with OUD and opioid misuse, respectively, from the NSDUH state reports, which follow the binomial data model below:
\begin{equation}
\label{eq:model-state-level-oud-and-misuse}
    \hat{Q}^{\text{OUD}}_s \mid \hat{Q}_s^{\text{misuse}}, r_s \sim \text{Binomial}\left(\hat{Q}_s^{\text{misuse}}, r_s\right).
\end{equation}
With the ratio $r_s$ properly defined, we can now link the prevalence of opioid misuse $p_i$ in the model with the estimated OUD prevalence $\hat{P}_i^{co}$ from the existing studies at the county level. Considering the nonnegativity of prevalence rates, we model the existing estimate $\hat{P}_i^{co}$ with a standard deviation $\hat{\sigma}_i^{co}$ for county $i\in\mathcal{I}_e$ of state $s_i\in\mathcal{S}$ following a lognormal distribution:
\begin{equation}
    \label{eq:model-county-level-OUD-prevalence}
    \hat{P}_i^{co} \mid p_i, r_{s_i}, \hat{\sigma}_i^{co} \sim \text{Lognormal}(u_i, v_i^2),
\end{equation}
where the distribution parameters $u_i$ and $v_i$ are defined as
\begin{subequations}\label{eq:lognormal-param-county-prev}
\begin{align}    
    u_i & = \log\left(p_i r_{s_i}\right) - v_i^2/2,\\
    v_i^2 & = \log\left(1+ (\hat{\sigma}_i^{co})^2/(p_i r_{s_i})^2\right),
\end{align}
\end{subequations}
which are derived from the method of moments by setting the mean of the lognormal distribution to be $p_i r_{s_i}$ and the standard deviation to be $\hat{\sigma}_i^{co}$. The standard deviation $\hat{\sigma}_i^{co}$ is calculated from the reported 95\% confidence interval (CI) for $\hat{P}_i^{co}$ by assuming $2\times 1.96\hat{\sigma}_i^{co}$ to equal the width of the CI.  

{\bf State-level estimates:} Denote $\hat{P}_s^{st}$ as the state-level estimate of opioid misuse in state $s\in\mathcal{S}$ from the NSDUH state reports. Let $\mathcal{I}_s$ denote the counties in state $s$. Similar to modeling the county-level estimates, we model the state-level estimator  $\hat{P}_s^{st}$ using a lognormal distribution,
\begin{equation}
\label{eq:model-state-level-prevalence}
\hat{P}_s^{st} \mid \{p_i\}_{i\in\mathcal{I}_s}, \{N_i\}_{i\in\mathcal{I}_s}, \gamma, \hat{\sigma}_{s}^{st} \sim \text{Lognormal}(\mu_s, \nu_s^2),
\end{equation}
where $\gamma$ is a multiplier to adjust for the underestimation ($\gamma\leq1$) from the NSDUH estimates, the location and scale parameters $\mu_s$ and $\nu_s$ are determined by matching the mean of the lognormal distribution to be $\bar{p}_s:=\gamma \frac{\sum_{i \in \mathcal{I}_s} p_i N_i}{\sum_{i \in \mathcal{I}_s} N_i}$ and the standard deviation to be $\hat{\sigma}_s^{st}$ with
\begin{subequations}\label{eq:lognormal-param-state-prev}
\begin{align}    
    \mu_s & = \log\left(\gamma \bar{p}_s\right) - \nu_s^2/2,\\
    \nu_s^2 & = \log\left(1+ (\hat{\sigma}_s^{st})^2 /\left(\gamma \bar{p}_s\right)^2\right).
\end{align}
\end{subequations}

Similar to $\hat{\sigma}_i^{co}$, the standard deviation $\hat{\sigma}_s^{st}$ is also calculated from the reported 95\% CI of state-level prevalence estimate $\hat{P}_s^{st}$ from NSDUH. All distributional assumptions are empirically verified and detailed results are presented later in Section \ref{sec:model diagnostics}.

{\bf State-level heterogeneity:} It has been commonly reported in the literature that not all communities are impacted by the opioid crisis equally across the country \citep{Monnat2018}. The above-defined baseline models \eqref{eq:p_oud}-\eqref{eq:q_death} capture differences in prevalence and overdose mortality rates across counties through a county's covariates, which, however, may not be sufficient to account for the differences across states due to a wide range of health, socioeconomic, and geographic factors (e.g., access to health services, unemployment, rurality) that could differ significantly by state \citep{Cano2023-cf}. Even states with comparable levels of naloxone distribution and access to treatment may exhibit different reductions in opioid overdose deaths \citep{Chhatwal2023}. Therefore, we propose incorporating state-level random effects into Equations \eqref{eq:p_oud} and \eqref{eq:q_death} to capture state-level heterogeneity not explained by the baseline models.

We perform a series of model diagnostics with residual analysis to determine the proper random effects that need to be added without potentially overfitting. We first examine the Pearson residuals of the estimated overdose deaths of each county from the baseline model without random effects and observe that the average residuals are significantly non-zero in over half of the states (Appendix Figure \ref{fig:appendix-residual-baseline-model}). This clearly indicates that between-state heterogeneity has not been adequately captured and thus warrants the addition of state-level random effects. Considering that the county-level overdose death information is relatively more complete in all states compared with the existing prevalence estimates, we start with adding the random effect to the intercept of the mortality rate \eqref{eq:q_death} to capture its heterogeneity across states. This addition substantially improves the model fit, with only 2 out of 42 states having the 95\% confidence intervals of state average residuals not covering zero (Appendix Figure \ref{fig:appendix-residual-rnd-intercpt-mortality-model}). Adding a random intercept for the prevalence rate \eqref{eq:p_oud} to the model results in only one state having an average residual significantly different from zero (Appendix Figure \ref{fig:appendix-residual-rnd-intercpt-prevalence-model}). We then regress the residuals on each county-level covariate, state by state, to determine whether the correlations of the residuals and the individual covariates vary substantially across states, such that adding random slopes may be warranted. Our results show that the correlation is not significant in more than 90\% of the states (Appendix Table
\ref{table:appendix-residual-regression}), indicating that random slopes are unnecessary. Based on these model diagnostic results, we modify Equations \eqref{eq:p_oud} and \eqref{eq:q_death} to the following:
\begin{align}
    \text{logit}(p_i)  & =
      \beta_0^p+b_{s_i, 0}^p + \bm{X}_i^{p}\bm{\beta}^{p}, & b_{s, 0}^p \sim   \mathcal{N}\left(0, (\sigma_{0}^p)^2\right),
    \label{eq:p_oud_RE} \\
    \text{logit}(m_i) & = 
      \beta_0^m+b_{s_i, 0}^m + \bm{X}_i^{m}\bm{\beta}^{m}, & b_{s, 0}^m \sim   \mathcal{N}\left(0, (\sigma_{0}^m)^2\right),
    \label{eq:q_death_RE}    
\end{align}
where the standard deviation $\sigma_0^m$ and $\sigma_0^p$ quantify the between-state heterogeneity and are model parameters to be estimated. 
Hereinafter, we proceed with the final model specification based on Equations \eqref{eq:model-mortality}-\eqref{eq:q_death_RE} for the remaining analyses throughout this paper.

\subsection{Likelihood and prior specifications} 
\label{sec:likelihood-and-priors}
The joint likelihood of our Bayesian hierarchical model $\mathscr{L}(\bm{\theta})$, given the model parameters $\bm{\theta}=(\beta_0^p, \bm{\beta}^p,  b_{s,0}^p, \sigma_0^p, \beta_0^m, \bm{\beta}^m, b_{s,0}^m, \sigma_0^m, \gamma, r_s)$, consists of three components corresponding to county-level overdose deaths, existing county- and state-level prevalence estimates. When calculating the likelihood of overdose deaths, we note that some observations $D_i$ in the publicly reported overdose death data are typically suppressed for privacy protection purposes if the actual count is less than or equal to a given threshold $c$ (e.g., $c=9$ for overdose death data from the public query of the CDC WONDER database). To capture this, we define the indicator $\delta_i=1$ if the actual count of overdose deaths is suppressed in county $i$ and $\delta_i=0$ otherwise. Then, the likelihood of overdose death data for county $i$ is given as:
\begin{equation}
\label{eq:likelihood-states-unknown}
\ell_i^D(\bm\theta) = \big[f_{pois}(D_i | m_i(\bm\theta) p_i (\bm\theta) N_i )\big]^{1-\delta_i} \big[F_{pois}(c | m_i(\bm\theta) p_i(\bm\theta) N_i )\big]^{\delta_i},
\end{equation}
where $f_{pois}(\cdot | m_i(\bm\theta) p_i (\bm\theta) N_i)\big)$ and $F_{pois}(\cdot | m_i(\bm\theta) p_i (\bm\theta) N_i)\big)$ are the probability mass function and cumulative distribution function of the Poisson distribution with expected value $m_i(\bm\theta) p_i (\bm\theta) N_i$, respectively. 

The likelihood functions for the data of existing prevalence estimates $\{\hat{P}_i^{co}\}_{i\in\mathcal{I}_e}$ at the county level and $\{\hat{P}_s^{st}, \hat{Q}_s^{\text{OUD}}\}_s$ at the state level follow directly from the model \eqref{eq:model-state-level-oud-and-misuse}, \eqref{eq:model-county-level-OUD-prevalence}, and \eqref{eq:model-state-level-prevalence}, with 
\begin{align}    
    \ell_i^{co} (\bm{\theta}) = & f_{lognormal}\left(\hat{P}_i^{co} \mid u_i(\bm{\theta}), v_i^2(\bm{\theta})\right), & \text{for } i\in \mathcal{I}_e, 
    \label{eq:likelihood-existing-prev-est-county}\\
    \ell_s^{st}(\bm{\theta}) = & f_{lognormal}\left(\hat{P}_s^{st} \mid \mu_s(\bm{\theta}), \nu_s^2(\bm{\theta})\right) \times f_{binomial} \left( \hat{Q}_s^{\text{OUD}} \mid \hat{Q}_s^{\text{misuse}}, r_s(\bm{\theta})\right), &\text{for } s\in\mathcal{S},
    \label{eq:likelihood-existing-prev-est-state}
\end{align}
where $f_{lognormal}(\cdot \mid u, v^2)$ is the probability density function for the lognormal distribution with location parameter $u$ and scale parameter $v$ on the log scale (see definitions in Equations \eqref{eq:lognormal-param-county-prev} and \eqref{eq:lognormal-param-state-prev}), and $f_{binomial}(\cdot \mid n, p)$ is the probability density function for the binomial distribution with $n$ trials and success probability $p$. Hence, the full likelihood is given as:

\begin{equation}
\label{eq:likelihood full}
  \mathscr{L}(\bm{\theta})
   =  
  \biggl(\prod_{i\in \mathcal{I}} \ell_i^D(\bm{\theta})\biggr)
   \times 
   \biggl(\prod_{i\in \mathcal{I}_e} \ell_i^{co}(\bm{\theta})\biggr)
   \times
  \biggl(\prod_{s\in \mathcal{S}} \ell_s^{st}(\bm{\theta})\biggr).
\end{equation}

We impose weakly informative priors for fixed effect intercepts $\beta_0^p, \beta_0^m\sim \mathcal{N}(0,10^2)$, the scalars and standard deviations $\gamma, r_s, \sigma_0^p, \sigma_0^m \sim \text{Half-Normal}(\sigma=10)$. For the fixed effect slopes $\bm{\beta}^p$ and $\bm{\beta}^m$, we apply the horseshoe+ prior to enable automatic shrinkage and selection of covariates \citep{horseshoe}. For each element $\beta_j$ of $(\bm{\beta}^p,\bm{\beta}^m)$,  its horseshoe+ prior is defined by the following hierarchical structure: 
\begin{align*} 
\beta_j \mid \lambda_j &\sim \mathcal{N}(0,\lambda_j^2), \\
\lambda_j \mid \tau, \zeta_j &\sim \text{Half-Cauchy}(0, \tau \zeta_j), \\
\zeta_j &\sim \text{Half-Cauchy}(0, 1),\\ 
\tau &\sim \text{Half-Cauchy}(0, 1),
\end{align*}
where $\lambda_j$ is the local shrinkage parameter for each individual component $\beta_j$, and $\tau$ serves as the global shrinkage parameter across multiple $\beta_j$. We define separate global shrinkage parameters for $\bm{\beta}^p$ and $\bm{\beta}^m$, respectively. As a sparsity-inducing prior, this hierarchical structure allows the model to adaptively shrink small coefficients towards zero while retaining larger coefficients that have a substantial impact, serving as a data-driven approach for variable selection. 

To examine the impact of priors on model performance, we evaluate the model with stronger priors in a series of sensitivity analyses. Specifically, we (1) change the prior for the fixed effect intercept $\beta_0^p, \beta_0^m$ to $N(0,5^2)$ and  $N(0,2.5^2)$, (2) change the prior for the random effect standard deviations $\sigma_0^p, \sigma_0^m$ to Half-Normal($\sigma$=5) and Half-Normal($\sigma$=2.5), and (3) replace heavy-tailed Half-Cauchy priors for the global and local shrinkage parameters of the horseshoe+ prior with stronger regularizing Half-Normal priors with the scale 0.5.

\subsection{Posterior Predictive Distribution}
\label{sec:predictive-interval}

We implement the Bayesian hierarchical model using the Stan programming language \citep{Gelman2015} and the \texttt{rstan} interface in R \citep{rstan}. Posterior samples are generated using the Hamiltonian Monte Carlo method \citep{HMC1, NUTS} with the No-U-Turn Sampler (NUTS), an efficient algorithm suited for high-dimensional and complex posterior distributions \citep{HMC}. All the computational studies in the following sections are performed on a MacBook Air laptop with an M2 chip processor and 16 GB RAM. We assess the model convergence based on the potential scale reduction factor $\widehat{R}<1.1$ \citep{Gelman1992}, and visually inspect the trace plots and autocorrelation functions to confirm that the parallel chains are well mixed and converge. We run NUTS in 4 parallel chains for the model fitting. For each chain, we collect samples from 50,000 iterations, with the first 25,000 iterations as the burn-in period and a thinning parameter of 10, resulting in a total number of 10,000 posterior samples across the 4 chains for estimating the posterior distribution of the model parameters.

To obtain the posterior predictive distribution for $\hat{P}_i^{co}$ and $\hat{P}_s^{st}$, we draw samples of $\{\hat{P}_i^{co,(k)}\}$ and $\{\hat{P}_s^{st,(k)}\}$ from the distributions \eqref{eq:model-county-level-OUD-prevalence} and \eqref{eq:model-state-level-prevalence}, respectively, given each posterior sample of $(p_i^{(k)}, \gamma^{(k)}, r_{s_i}^{(k)})$. Similarly, posterior predictive samples of overdose deaths, $\{D_i^{(k)}\}$, are drawn from the Poisson distribution following \eqref{eq:model-mortality} given $N_i$ and posterior samples $(p_i^{(k)}, m_i^{(k)})$. The 95\% prediction interval of posterior samples is defined based on their 2.5-th and 97.5-th percentiles.

\subsection{Model diagnostics and validation}
\label{sec:model diagnostics}

We first inspect the residuals to check the compatibility of distributional assumptions. Specifically, we calculate the residuals between the observed values and their model-predicted counterparts based on the posterior mean of parameters for (1) overdose deaths (unsuppressed), (2) county-level prevalence estimates, and (3) state-level prevalence estimates, respectively. Given the assumed Poisson distribution of overdose deaths, we obtain Pearson residuals for overdose deaths (unsuppressed) by dividing the difference between the observed value and the expected value by the square root of the variance of the expected value. 

The posterior predictive checks compare the posterior distribution of total overdose deaths aggregated at the state level with the observed overdose deaths by each state. New queries are made to the CDC WONDER Multiple Cause of Deaths database for the state-level number of opioid overdose deaths, which does not suffer from suppressed data. For counties with suppressed values, we calculate the probability that the posterior predictive distribution of $D_i$ falls below the suppression threshold $c$.  

In addition, we perform 10-fold cross-validation (CV) to evaluate the model's capability in prediction on data that have not been seen. For counties in $\mathcal{I}_e$ with existing OUD prevalence estimates, we randomly split these counties into 10 folds, holding out the $\hat{P}_i^{co}$ data in one fold at a time to fit the model, and then compare the mean and prediction intervals from the posterior samples with the holdout observations. For the overdose outcome, we conduct a stratified 10-fold cross-validation for all counties $\mathcal{I}$, with counties in $\mathcal{I}_e$ randomly distributed across all folds. We hold out
overdose deaths for one fold at a time to fit the model and obtain the posterior samples of the expected overdose deaths $m_i^{(k)} p_i^{(k)} N_i$ given the posterior samples $(p_i^{(k)}, m_i^{(k)})$. We evaluate the mean absolute percentage error (MAPE), percentage of overlapping interval estimates, and percentage of our interval estimates covering the observations of prevalence $\{\hat{P}_i^{co}\}_{i\in\mathcal{I}_e}$ and the unsuppressed overdose deaths $\{D_i: \delta_i=0, i\in\mathcal{I}\}$. For the counties with suppressed overdose deaths, we report the percentage of our prediction intervals that overlap with the range of suppressed data (i.e., $\le 9$).

To further explore the model's generalizability beyond the dependence of data from state-specific studies, we also cross-validate the model by holding out the observed OUD prevalence data $\hat{P}_i^{co}$ by state. Since the county-level OUD prevalence data included in our real-world study are from only two states (see Section \ref{sec:real-world data settings} for more detailed data descriptions), holding out the data from one state leaves the model only with the observed prevalence $\hat{P}_i^{co}$ from the other state to be fitted from. In our exploratory analysis, we find that such data sparsity renders the full model \eqref{eq:model-mortality}-\eqref{eq:q_death_RE} unidentifiable given the multiple random effects in the current model; therefore, we remove the random intercept in the prevalence rate \eqref{eq:p_oud_RE} to restore the model identifiability and use the modified model within this leave-one-state-out analysis.

\section{Results}
\label{sec:results}
\subsection{Real-world data settings}
\label{sec:real-world data settings}
In this section, we apply our proposed Bayesian hierarchical model to real-world data to estimate county-level prevalence of opioid misuse in all 50 states in the US (excluding Washington DC). Our analysis uses 2015 as a reference year to be consistent with a key study of estimating county-level OUD prevalence estimates in Massachusetts using capture-recapture analysis \citep{Barocas2018-ke}, while drawing on additional data sources from the closest available years to maintain temporal proximity. For existing county-level estimates $\{\hat{P}_i^{co}\}_{i\in\mathcal{I}_e}$, in addition to those in the 2015 Massachusetts study \citep{Barocas2018-ke}, we also include the estimates from Kentucky in 2017 \citep{Thompson2023}. Despite the two-year difference, including county-level prevalence data from more than one state is crucial to allow the model to properly identify the random effects designed for capturing state heterogeneity. To account for the year difference, we allow the standard deviation of the reported county-level prevalence estimates in Kentucky to be an unknown parameter $\sigma_{KY}^{co}$, to be inferred from the data, and use it to replace the $\hat{\sigma}_i^{co}$ as calculated directly from the reported interval estimates.

For the state-level prevalence estimates $\{\hat{P}_s^{st}\}_{s\in\mathcal{S}}$, we combine the state-level estimates of pain reliever use disorder and heroin use from the NSDUH 2015-2016 state report to represent the opioid misuse prevalence in 2015. In addition, the NSDUH 2021-2022 state report provides the earliest estimates of both OUD and opioid misuse at the state level, allowing us to utilize these data to infer the model parameter $r_s$ by state. 

The number of opioid overdose deaths $\{D_i\}$ for all counties in 2015 are queried from the CDC WONDER database. We used the 10-th revision of the International Classification of Diseases (ICD-10) underlying cause of deaths (UCD) codes for unintentional (X40-X44), suicide (X60-64), homicide (X85), or undetermined (Y10-Y14), and the multiple causes of deaths (MCD) codes for heroin (T40.1), natural and semisynthetic opioids (T40.2), Methadone (T40.3), synthetic opioids other than methadone (T40.4), and other and unspecified narcotics (T40.6), following the commonly used selection criteria in the literature \citep{FeuersteinSimon2020, Seth2018}. 

We include the following covariates as potential prevalence and mortality determinants considering their broad relevance and potential influences to opioid and substance use: (1) the annual opioid dispensing rate (per 100 population), to capture the supply-side effect of the opioid epidemic \citep{Vuolo2022, Thombs2020}, (2) adult smoking rate, a risk factor that is strongly linked to other health-related risk behaviors and could serve as a proxy for broader substance use behaviors \citep{Zale2014, YoungWolff2017}, (3) some college education rate (the percentage of adults aged 25 and over who have at least completed one year of college education), as educational attainment could influence risk behaviors and access to health resources \citep{Powell2023, Thombs2020}, (4) unemployment rate, considering that lack of employment and financial stress could lead to increased substance misuse and higher overdose risks \citep{Rudolph2020, Saba2025}, and (5) number of primary care physicians per 100,000 population as a measure of access to healthcare and basic health needs \citep{Haffajee2019, Eichmeyer2023}. Opioid dispensing rates in 2015 are obtained from the CDC based on a retail prescription dispensing database \citep{CDCDispensingRates}, and other covariates of county socio-demographic factors are retrieved from the County Health Rankings in 2015 \citep{countyhealth2024}. Our full data set consists of 3,132 counties from 50 US states. Only $<$0.5\% of the counties have missing values in covariates, which are imputed by the state mean. All variables are standardized prior to the model fitting.

\subsection{Model goodness-of-fit}
Our model diagnostics verify that our proposed hierarchical model shows reasonable goodness of fit to the given data (see Appendix Figure \ref{fig:distributional-assumptions}). The residuals appear to be centered around zero without a strong correlation with the fitted values, and the expected values predicted by the model are aligned with observed values for all three observed outcomes, implying reasonable distributional assumptions in our model. 

Our posterior predictive checks further show that the observed numbers of the state-level overdose deaths fall within the high probability regions based on our posterior predictive distribution in most of the states (see Appendix Figure \ref{fig:ppc}), which also suggests that our proposed model has reasonable assumptions and is properly fitted. We also find that in some states, the reported overdose death numbers fall on the lower end of the predictive distribution, and these states tend to have higher prevalence or have counties with high prevalence estimates. Such under-representation of the reported overdose deaths compared with the model predictive samples may be attributed to the high proportion of counties with suppressed data in these states. As the inference of model parameters draws more information from counties with unsuppressed death data ($\ge 10$) than those with suppressed data, applying these fitted parameters to the suppressed counties tends to place the posterior distribution towards the suppression threshold, resulting in predicted values higher than reported observations.

\begin{figure}[ht!]
\centering
\includegraphics[width=0.95\linewidth]{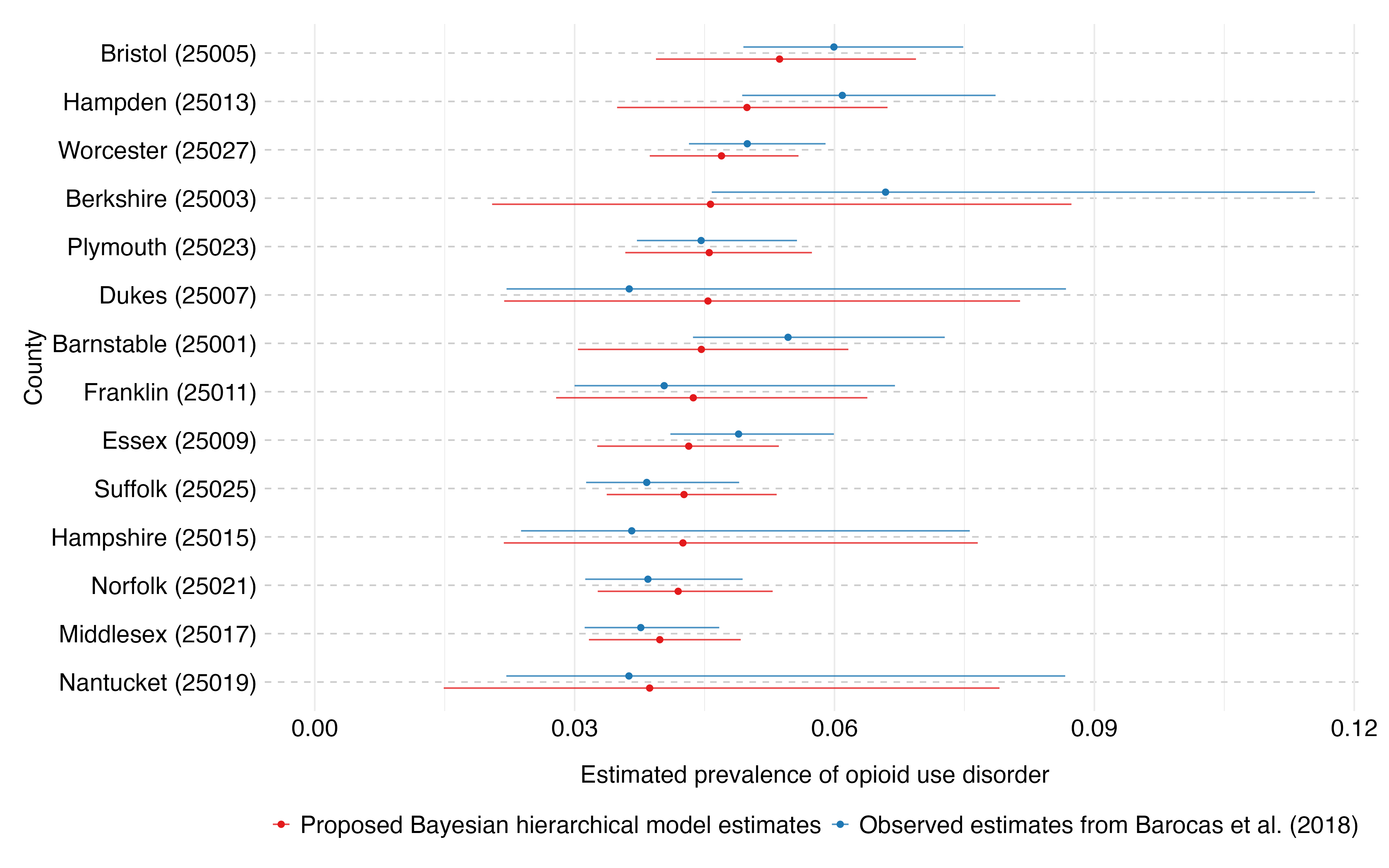}
\caption{The 10-fold cross-validation results of the Bayesian hierarchical model compared with the estimates from \citet{Barocas2018-ke}.}
\label{fig:10foldcv-ma}
\end{figure}

To further validate our model's county-level prevalence estimates, we compare our results with existing estimates in Massachusetts \citep{Barocas2018-ke} and Kentucky \citep{Thompson2023} via 10-fold cross-validation. Overall, our model achieves a MAPE of 45.2\% and a Spearman's rank correlation coefficient of 0.60 among all 134 counties in the two states combined. Our results show higher accuracy for counties in Massachusetts with a MAPE of 12.8\% and a correlation coefficient of 0.78 (Figure \ref{fig:10foldcv-ma}), compared with Kentucky with a MAPE of 49.0\% and a correlation coefficient of 0.53 (Appendix Figure \ref{fig:10foldcv-ky}). Moreover, the 95\% prediction intervals from our model cover observed estimates in 95.5\% (128 out of 134) counties in the two states, supporting the practical validity and relevance of the estimates from our model. 

We also perform the 10-fold cross-validation for the overdose death outcome, analyzed separately for counties with unsuppressed and suppressed overdose death data. Among counties with unsuppressed death counts, the observed numbers of overdose deaths in 63.3\% (389 out of 615) counties are covered by our model's 95\% prediction interval, resulting in a MAPE of 37.6\% for the point estimates. For those counties with suppressed data (i.e., observed deaths $\le 9$), the estimated overdose deaths by our model fall below the suppression threshold $c=9$ in 90.8\% (2,286 out of 2,517) counties, and the probability that the posterior predictive distribution of overdose deaths falls below the threshold $c$ has an average of 0.91 across these counties and is $\ge$0.95 in 79.1\% (1,992 out of 2,517) counties.

\subsection{County-level prevalence estimates}

Using the real-world data from 3,132 counties of 50 US states, our model estimates the overall opioid misuse prevalence ($\sum_i p_i N_i/\sum_i N_i$) of 5.1\% (95\% CrI: 3.7\%, 6.8\%), equivalent to 14.10 (10.22, 18.66) million individuals with opioid misuse in 2015 among the population aged 11 years and above in the US. At the county level, the estimated prevalence of opioid misuse ($p_i$) has a median of 5.0\%, with the first and third quartiles of 4.3\% and 5.9\%, respectively. There are 1,672 (53.4\%) counties having an estimated prevalence below the national average. The histogram and the distribution for prevalence estimates of all counties are shown in Appendix Figure \ref{fig:histogram}. 

Figure \ref{fig:results-map} shows the posterior predictive mean (Figure \ref{fig:map-prevalence-estimate}) and standard deviation (Figure \ref{fig:map-prev-sd}) of opioid misuse prevalence estimates across all counties. Counties with the highest 1\% standard deviation are not shown in the maps, considering their high uncertainty. As expected, estimates in the counties of Massachusetts and Kentucky have the lowest uncertainty (as indicated by the blue regions with low posterior standard deviations in Figure \ref{fig:map-prev-sd}), because these two states are supplied with county-level OUD prevalence data, which have effectively concentrated their posterior estimates around the data and lead to high confidence. Counties with high prevalence are concentrated in the Appalachian region, including northeastern Kentucky and Ohio, large parts of West Virginia, and Tennessee, as well as New England and the Pacific West regions, with notably higher uncertainty associated with their estimates than those in Kentucky and Massachusetts. As the high prevalence estimates in these areas are mainly driven by the county-level covariates, the model also propagates the variance of the covariate coefficients to high uncertainty of the prevalence estimates in the absence of direct county-level prevalence data. The state-level prevalence estimates are shown in Appendix Figure \ref{fig:prevalence-est-state-level}. The complete results of the county-level prevalence estimates are provided in the supplemental file. 

\begin{figure}[h!]
  \centering
  \begin{subfigure}[h]{\textwidth}
    \includegraphics[width=\textwidth]{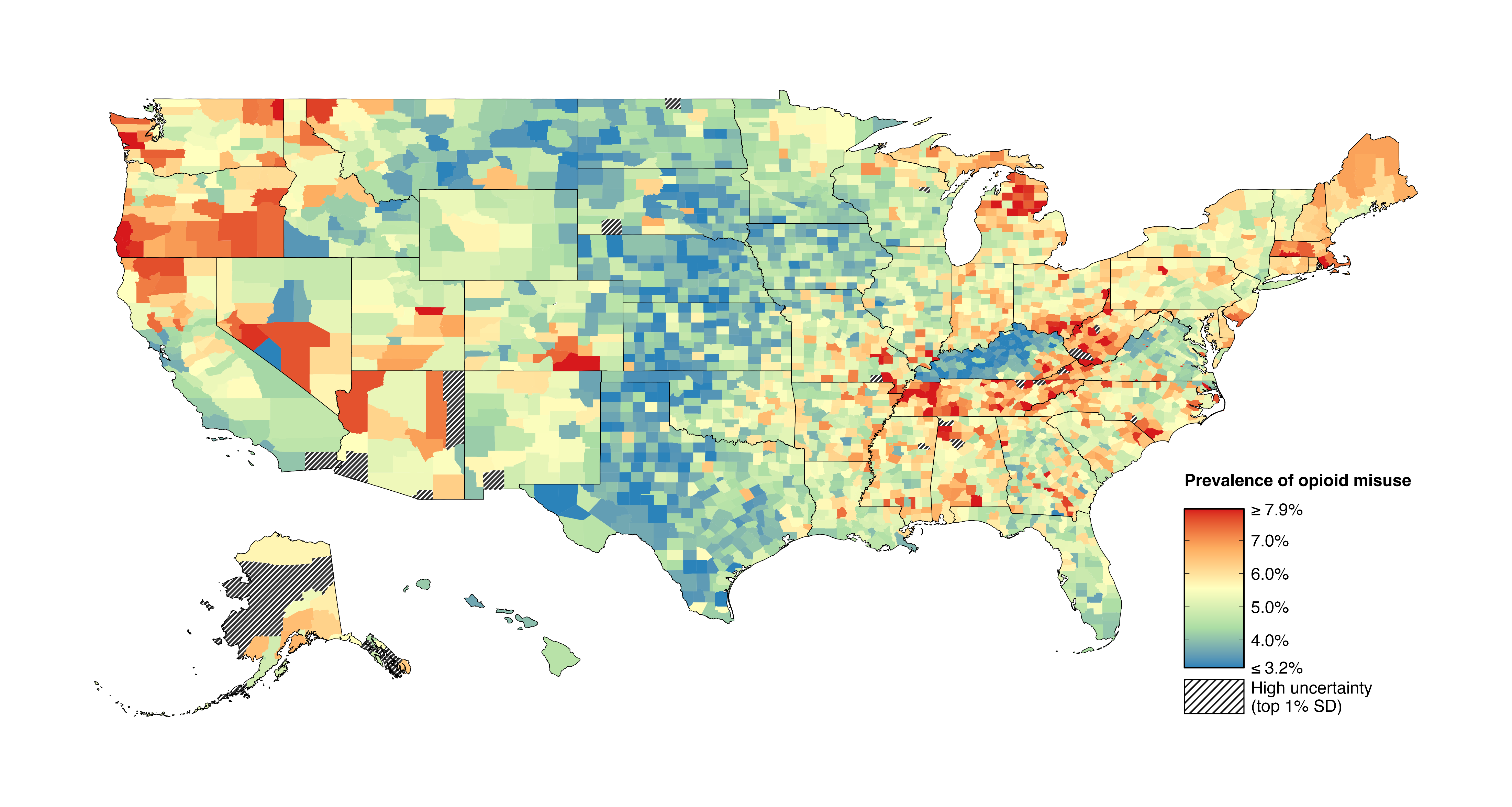}
    \caption{Estimated prevalence of opioid misuse in all US counties among the population aged 11 years and above in 2015.}
    \label{fig:map-prevalence-estimate}
  \end{subfigure}
    \begin{subfigure}[h]{\textwidth}
    \includegraphics[width=\linewidth]{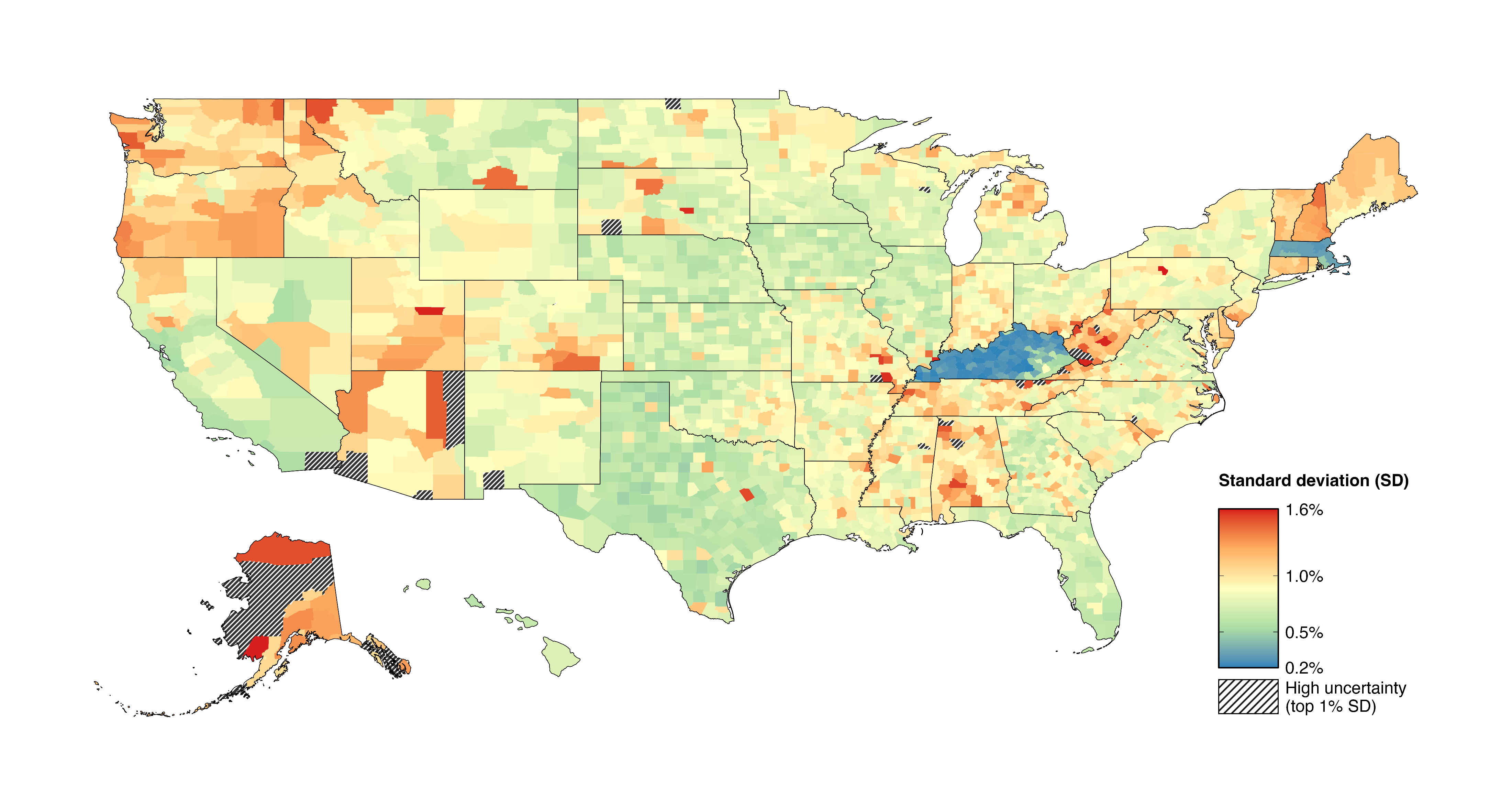}
    \caption{Standard deviation of estimated prevalence of opioid misuse in all US counties.}
    \label{fig:map-prev-sd}
    \end{subfigure}
  \caption{Results of county-level estimate of opioid misuse prevalence in the US.}
  \label{fig:results-map}
\end{figure}

Estimates of model parameters are shown in Table \ref{tab:parameter-est-values}. For the {\em prevalence rate}, the fixed-effect intercept ($\beta_0^p$) is estimated to be -2.90 (95\% CrI: -3.09, -2.70), translating to a nationwide baseline prevalence level of 5.2\% (4.4\%, 6.3\%). Among the county-level covariates, the opioid dispensing rate shows a significant positive association with the prevalence rate (with $\hat{\beta}_1^p=$ 0.16 [0.09, 0.22]), which is in line with the findings regarding the role of opioid dispensing in the opioid epidemic from the literature \citep{Cano2023-cf,monnat2019using}. Unemployment rate is also shown to be positively associated with the prevalence (with $\hat{\beta}_4^p=$ 0.13 [0.05, 0.20]), suggesting the potential link between economic instability and higher prevalence of opioid misuse \citep{Azagba2021-zx}. Other covariates do not show significant effects, with their coefficients being shrunk close to 0 due to the horseshoe+ prior (see Section \ref{sec:likelihood-and-priors}). For the {\em mortality rate}, the covariate of adult smoking rate exhibits a significant positive association with the overdose mortality rate in the county (with $\hat{\beta}_2^m=$ 0.15 [0.07, 0.23]), which is consistent with the previous literature that has shown smoking rate positively associated with opioid misuse \citep{rajabi2019association} and increased overdose mortality \citep{langabeer2022county}, potentially through complex interactions with other social and behavioral factors. We observe that unemployment rate is negatively associated with the mortality rate (conditioned on the opioid misuse), implying that its overall association with the total number of overdose deaths may be unclear, consistent with the literature showing that these associations vary across regions and subpopulations \citep{kerry2019spatial, hollingsworth2017macroeconomic,frankenfeld2019county}.

Importantly, the random effects in both prevalence and mortality rates have successfully captured the heterogeneity across states as intended, with significantly non-zero estimates of the standard deviation of the random effect intercepts $\sigma_0^m=0.48$ (0.38, 0.61) and $\sigma_0^p=0.21$ (0.14, 0.28). Furthermore, our model adjusts for the underestimation of the NSDUH estimates by the estimated $\gamma=$ 0.21 (0.17, 0.24), which implies an approximate multiplier of 4.9, on average, applied to the NSDUH estimates. Due to the state-level random effects in our model, the actual relative scales of our estimates compared with the NSDUH state-level prevalence estimates vary by state between 0.15 to 0.36 (Appendix Figure \ref{fig:cmp-nsduh-state-level}). These scales correspond to multipliers ranging 2.8-6.7, which are similar to the ranges of the reported multipliers between 2.9-8.0 from the multiplier method studies that use NSDUH as the benchmark data source \citep{Mojtabai2022,lim2024improving,Keyes2022-zv}.

\begin{table}[h!]
\caption{Posterior means and 95\% credible intervals of model parameters.}
\label{tab:parameter-est-values}
\small
\centering
\begin{tabular}{@{}lcc@{}}
\toprule
\textbf{\begin{tabular}[c]{@{}c@{}}Parameter \\ \end{tabular}} & \textbf{\begin{tabular}[c]{@{}c@{}}Posterior \\ mean\end{tabular}} & \textbf{\begin{tabular}[c]{@{}c@{}}95\% Credible \\ Interval\end{tabular}} \\ \midrule
\textbf{Prevalence rate} &  &  \\
\hspace{1em} Fixed-effect intercept ($\beta_0^p$) & -2.90 & (-3.09, -2.70)* \\
\hspace{1em} Coefficient for opioid dispensing ($\beta_1^p$)& 0.16 & (0.09, 0.22)* \\
\hspace{1em} Coefficient for adult smoking ($\beta_2^p$) & 0.08 & (-0.00, 0.16) \\
\hspace{1em} Coefficient for some college education ($\beta_3^p$) & 0.05 & (-0.02, 0.12) \\
\hspace{1em} Coefficient for unemployment ($\beta_4^p$) & 0.13 & (0.05, 0.20)* \\
\hspace{1em} Coefficient for primary care physicians ($\beta_5^p$) & 0.01 & (-0.03, 0.04) \\
\hspace{1em} Standard deviation of the random-effect intercept ($\sigma_0^p$) & 0.21 & (0.14, 0.28)* \\
\textbf{Mortality rate} &  &  \\
\hspace{1em} Fixed-effect intercept ($\beta_0^m$) & -6.01 & (-6.23, -5.78)* \\
\hspace{1em} Coefficient for opioid dispensing ($\beta_1^m$)& 0.06 & (-0.00, 0.13) \\
\hspace{1em} Coefficient for adult smoking ($\beta_2^m$) & 0.15 & (0.07, 0.23)* \\
\hspace{1em} Coefficient for some college education ($\beta_3^m$) & 0.04 & (-0.02, 0.12) \\
\hspace{1em} Coefficient for unemployment ($\beta_4^m$) & -0.16 & (-0.23, -0.07)* \\
\hspace{1em} Coefficient for primary care physicians ($\beta_5^m$) & 0.00 & (-0.03, 0.03) \\
\hspace{1em} Standard deviation of the random-effect intercept ($\sigma_0^m$) & 0.48 & (0.38, 0.61)* \\
\textbf{Existing prevalence estimates} &  &  \\
\hspace{1em} Scaling coefficient for NSDUH estimates ($\gamma$) & 0.21 & (0.17, 0.24)* \\ 
\hspace{1em} Proportion of OUD among opioid misuse in MA ($r_{MA}$) & 0.64 & (0.63, 0.64)* \\
\hspace{1em} Proportion of OUD among opioid misuse in KY ($r_{KY}$) & 0.96 & (0.95, 0.96)* \\
\hspace{1em} Estimated standard deviation of KY estimates ($\sigma_{KY}$) & 0.02 & (0.01, 0.02)* \\ \bottomrule
\multicolumn{3}{l}{\footnotesize * The 95\% credible interval does not cover zero, indicating statistical significance.}
\end{tabular}
\end{table}

Lastly, to further validate the practical relevance of our results, we cross-compare them with similar estimates from other state-specific studies of estimating county-level opioid misuse prevalence. A relevant study for comparison is a Bayesian hierarchical modeling analysis for estimating the prevalence of opioid misuse in North Carolina during 2016-2021 \citep{kline2025estimating}. Acknowledging the differences in study design and setting, we observe that our estimated prevalence rates tend to be higher than those of 2016 from Kline's study. Still, the overall trend is consistent, with a significant correlation of 0.50 (95\% CrI: 0.34-0.64) between the two sets of estimates. Our 95\% CrIs cover their point estimates in 61 of 100 counties and overlap with their interval estimates in 97 of 100 counties. We also compare with an Ohio-based study that estimates the OUD prevalence among the Medicaid population in 2019 \citep{Doogan2022-lf}. Although our estimates are lower, likely due to that the Medicaid population tend to have a higher OUD prevalence than the general population, the Spearman's rank correlation between the two sets of estimates remains high at 0.66 (0.29-0.86), indicating consistency in terms of their relative sizes. Several other state-specific studies in county-level opioid misuse or use disorder prevalence estimation (e.g., for New York \citep{SantaellaTenorio2024} and Ohio \citep{Hepler2023-mu}) do not publish numeric values of their county-level estimates and thus are not included for the comparison. 

\subsection{Sensitivity analysis}
To further evaluate the model's prediction capability beyond the states with county-level prevalence data, we perform leave-one-state-out cross-validations by holding out the OUD prevalence estimates of one state (i.e., Massachusetts or Kentucky) at a time. Our model recovers the rank order of prevalence estimates in the counties of the held-out state, demonstrating the model's prediction capability beyond the states with county-level prevalence estimates. In particular, the OUD prevalence estimated by the KY-based model (with the OUD prevalence data held out in Massachusetts counties) show a strong positive correlation with the actual prevalence data in Massachusetts (with Spearman’s correlation $\rho = 0.75$ and Pearson’s correlation $r = 0.75$, p-value $< 0.05$), and similarly for the prevalence estimates in Kentucky by the MA-based model ($\rho = 0.47$ and $r = 0.61$, p-value $< 0.05$). On the other hand, since these single-state-based models do not have state-level random effects to capture between-state heterogeneity, the absolute values of the prevalence estimates from these models have limited generalizability. For example, the MA-based model estimates the national prevalence to be 8.25\% (95\% CrI: 6.98\%, 9.69\%), substantially higher than 5.1\% (3.7\%, 6.8\%) estimated by the full model, whereas the KY-based model yields a much lower estimate of 2.78\% (95\% CrI: 2.32\%, 3.31\%). Similarly, at the county level, the MA-based model consistently overestimates and the KY-based model underestimates the prevalence compared with our full model that strikes a balance in between (Appendix Figure \ref{fig:loso_9states}). These discrepancies also underscore the importance of capturing cross-state variations for estimating the prevalence of opioid misuse and the need for incorporating county-level OUD prevalence from multiple states as references, which is implemented in our full model.

Our sensitivity analysis on the prior shows that the model results are highly robust to the prior specifications of our Bayesian model. With different prior specifications, the computational time remains at a similar level, with approximately 1200-1400 seconds per 5000 iterations in the \texttt{Stan} software. All models converge with the maximum $\hat{R}$ of 1.08. As illustrated in Appendix Figure \ref{fig:prior_sens}, the estimates of all model parameters remain consistent across all alternative settings with more strict priors compared with the base case, confirming that the inference results are primarily driven by the information from the data (with all covariates being standardized) rather than the scale of the priors.

\section{Discussion}
\label{sec: discussion}
In this study, we develop a multi-state Bayesian hierarchical model that synthesizes publicly available data to estimate the county-level prevalence of opioid misuse nationwide in the US. Cross-validation shows high prediction accuracy of our proposed model compared to existing prevalence estimates and observed overdose deaths. Our study contributes to the literature on county-level prevalence estimation for opioid and substance misuse by presenting a unified framework that seamlessly integrates data currently available to all counties, without relying on state-specific {\em ad-hoc} data sources. 

Although several states have estimated county-level prevalence of opioid misuse or OUD using state-specific data sources, not all states are equipped with similar comprehensive data resources to support such analyses; hence, the implications of these state-specific models and corresponding results on other states are limited. In contrast, our model is designed to be broadly applicable to all states and only requires publicly and widely available information, which alleviates the burden of obtaining and preparing data in states where data infrastructure is lacking. To the best of our knowledge, we are the first to jointly model county-level opioid misuse for all 50 US states. Our results serve as an initial step towards better understanding the local burden of the opioid crisis and helping to inform effective resource allocation strategies to mitigate the opioid crisis.

A recent study by \cite{Peterson2025} has pursued a similar objective of estimating county-level OUD prevalence nationwide using a Bayesian modeling framework, but our model has several important distinctions. Peterson's spatio-temporal model adopts a top-down modeling strategy, which starts with state-level prevalence estimates derived from NSDUH and distributes the state totals into each county, using overdose mortality data as one of the covariates. In contrast, our data approach integrates state-level prevalence data from NSDUH (for all states), county-level prevalence estimates from epidemiological studies (for limited states), and overdose mortality data into a unified likelihood. In addition, we model the overdose deaths as an observed outcome of opioid misuse, enabling principled handling of suppressed counts and incorporating the uncertainty in overdose death data. There is also a subtle but important difference in the definition of the target population between these two studies. \citet{Peterson2025} focus on the OUD population while addressing the changing diagnostic criteria in NSDUH over time, whereas our work estimates the prevalence for a broader opioid misuse population at risk of overdose. Our model also explicitly formulates the link between these two concepts of OUD and opioid misuse, enabling the flexibility of integrating prevalence estimates based on different definitions from multiple studies.  

In this study, we utilize the publicly available overdose death data from a national data source to enable the model's wide applicability across the country, but it is important to acknowledge the underlying heterogeneity in such a data source, which leads to a study limitation. While our model only considers a state-level random effect to capture the heterogeneity at a high level, the overdose death data are subject to the varying toxicity of the drug supply in local illicit drug markets (e.g., the penetration of fentanyl) that could vary substantially across regions and even within states \citep{ciccarone2019triple}. Rural-urban disparities could also contribute to the county-level heterogeneity in overdose mortality outcomes due to the differences in available resources for overdose responses. In addition, the accuracy of the documented drug types in overdose deaths is subject to the reporting practice of the coroner's office, which may vary substantially across states and counties. Evidence from previous studies has also shown the substantial under-reporting and misclassifications of opioid overdose deaths in the vital statistics data \citep{ruhm2016drug}. Due to data limitations, our model cannot fully capture all the possible sources of heterogeneity in overdose death data, which might have contributed to the discrepancies observed for some states in our posterior predictive checks. Nevertheless, the cross-validation results have shown a reasonable predictive accuracy on average, demonstrating our model's ability to draw useful information from nationwide publicly available data to address the prevalence estimation problem.

Another limitation of the current study is its cross-sectional design, coupled with data sources that are not drawn from the exact same year. While we adopt 2015 as the reference year for our analysis based on a seminal study of estimating county-level OUD prevalence \citep{Barocas2018-ke}, other data sources are selected to be as close as possible, subject to data availability. Considering that the county-level prevalence estimates for Kentucky in 2017 are used as model input to inform inference in a different year, we assume a higher uncertainty of these estimates by treating their standard deviation as a model parameter to be inferred, instead of fixing it based on the reported values directly. To inform the state-specific relative ratio between OUD and opioid misuse, $r_s$, we rely on the estimates $\hat{Q}_s^{\text{misuse}}$ and $\hat{Q}_s^{\text{OUD}}$ that are only available in 2021-2022 NSDUH state report or later, implicitly assuming the temporal stability of these ratios. Explicit modeling of the temporal dynamics would provide a principled approach to integrating data sources from different time points. As additional studies emerge, new data are more likely to have improved temporal alignment, which may also mitigate the limitation arising from the time differences in our current model.

Our proposed approach for public data integration should not replace the efforts of state-specific studies that focus on estimating the county-level prevalence of opioid misuse using ad-hoc datasets available within the state. In fact, our approach can be further enhanced by leveraging more estimates published from state-specific studies in the future. In the case study presented here, we use the existing county-level estimates for OUD prevalence in Massachusetts and Kentucky, which have been systematically estimated from the capture-recapture analyses based on multiple linked datasets within each state. If other states can perform the capture-capture analysis with a similar design, their estimates can be readily incorporated into our current model. Moreover, our approach can be extended to accommodate state-specific estimates derived from other statistical methods (e.g., integrated abundance model, multiplier method) and for different target populations (e.g., opioid misuse vs. OUD), as such estimates become available in the growing literature. For example, separate models could be constructed with modifications to align with the estimation approach used in each state-specific study, and their inference results could be synthesized through an ensemble framework. The integration of existing estimates, which are available but siloed across different state-specific studies, allows us not only to triangulate the estimates from different approaches but also to maximize their value to inform and refine the overall understanding of opioid use and misuse in other states and counties.

Our study lends itself to several directions for future research. A promising extension of our current model is to incorporate spatial correlations between counties to account for geographic and social proximity, allowing the model to borrow information from other counties and potentially help reduce across-county variations in the estimates. Although powerful, spatial models may suffer from computational challenges, especially at the nationwide county-level scale, and thus may warrant further development and refinement of computational approaches. Another direction of expanding our current model without substantial structural changes is to incorporate additional relevant social, demographic, and emerging epidemiological factors as covariates in the model, which could then be systematically screened by feature selection approaches, to help better capture heterogeneity across counties and states. The model could also be further refined by adjusting for the underestimation of opioid involvement in overdose deaths due to missing data on drug specificity in toxicology reports \citep{gutkind2024misclassification}. While the completeness of drug classification information has improved in more recent years \citep{boslett2019unclassified}, drug specificity in death records has become increasingly important, considering the poly-substance use and emerging new drug threats in the current stage of opioid crisis. Lastly, it may also be of policymakers' interests to further assess the contributions of different data sources and potential value of acquiring additional data, for example, through the value of information analysis \citep{keisler2014value,parsons2022unified}, which could help guide policymakers in identifying priority areas for additional data collection efforts to improve the estimation of opioid misuse prevalence across the country. 

\section*{Acknowledgments}
This work was supported by the National Science Foundation (NSF) under grant agreement CMMI-2240408, the National Institutes of Health (NIH)/National Institute of Allergy and Infectious Diseases (NIAID) under award numbers R01AI136664 and R01AI170249. Any opinions, findings and conclusions or recommendations expressed in this material are those of the author(s) and do not necessarily reflect the views of the NSF or NIH.

\bibliographystyle{apalike}
\bibliography{mybib.bib} 

\clearpage
\newpage
\appendix
\renewcommand\thesection{\Alph{section}}

\setcounter{table}{0}
\setcounter{figure}{0}

\counterwithin{table}{section}
\counterwithin{figure}{section}

\section{Appendix: Supplementary Results}

\begin{figure}[ht!]
\centering
\includegraphics[width=0.8\linewidth]{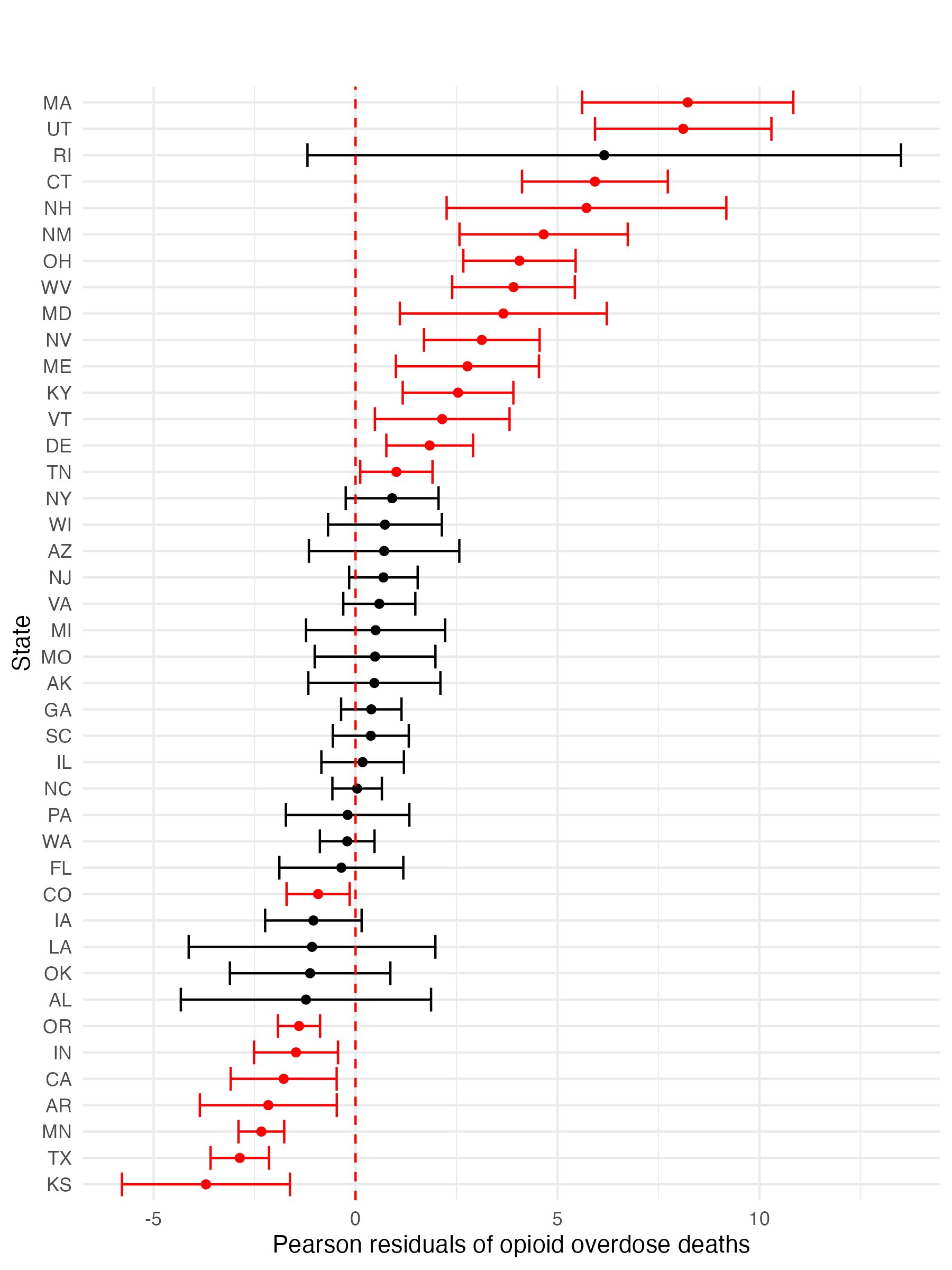}
\caption{Pearson residuals of opioid overdose deaths in the baseline model for counties with unsuppressed observations by state. The point represents the mean residual in the given state, and the line indicates the corresponding 95\% confidence interval (CI). More than half of the states (22 out of 42 states that have at least 3 counties with unsuppressed overdose deaths) have significantly non-zero residuals as shown by the CIs highlighted in red, suggesting that between-state heterogeneity has not been adequately captured in this model.}
\label{fig:appendix-residual-baseline-model}
\end{figure}

\begin{figure}[ht!]
\centering
\includegraphics[width=0.8\linewidth]{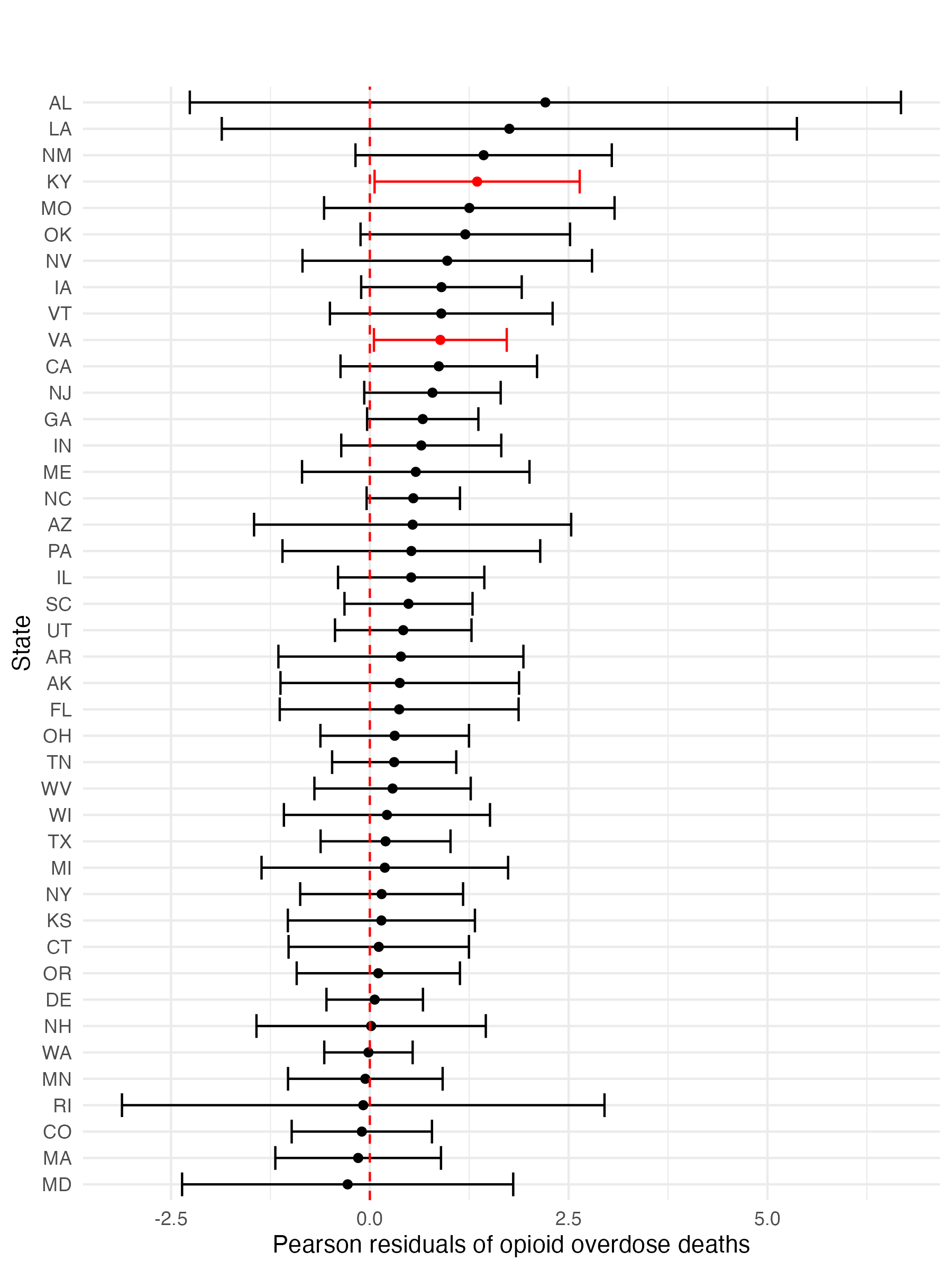}
\caption{Pearson residuals of opioid overdose deaths after adding a random intercept to the mortality rate in the model for counties with unsuppressed observations by state. The point represents the mean residual in the given state, and the line indicates the corresponding 95\% confidence interval. For the majority of the states (40 out of 42 states that have at least 3 counties with unsuppressed overdose deaths), the average residual is not significantly different from zero.}
\label{fig:appendix-residual-rnd-intercpt-mortality-model}
\end{figure}

\begin{figure}[ht!]
\centering
\includegraphics[width=0.8\linewidth]{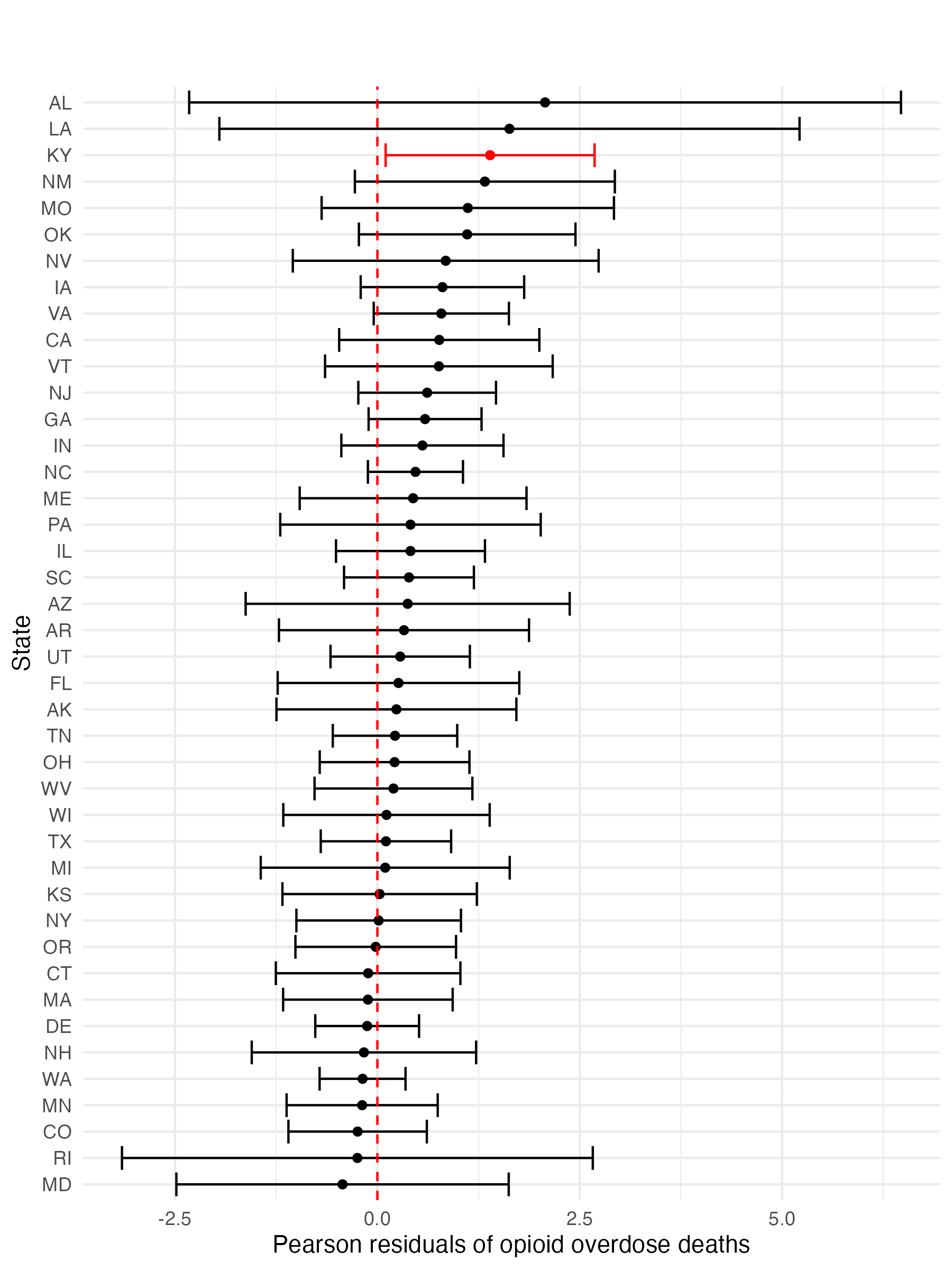}
\caption{Pearson residuals of opioid overdose deaths by state after adding random intercepts to mortality and prevalence rates in the model. The point represents the mean residual in the given state, and the line indicates the corresponding 95\% confidence interval. Only one state shows that the average residual is significantly different from zero. }
\label{fig:appendix-residual-rnd-intercpt-prevalence-model}
\end{figure}

\begin{table}[ht!]
\centering
\caption{Results of univariate regression analyses for each state by regressing the residuals of overdose deaths on the covariate of each county, after adding state-level random intercepts to the mortality and prevalence rates in the Bayesian hierarchical model. This table summarizes the number of states with significant intercepts and slopes from the univariate regression (among the 42 states with at least 3 counties having unsuppressed overdose death data). These results show that the residuals of the current model remain significantly correlated with the covariates in only a few states, implying that adding a random slope to any of these covariates in the model is deemed unnecessary.}
\label{table:appendix-residual-regression}
\begin{tabular}{lcc}
\toprule
\textbf{Covariate} & \textbf{Significant intercepts} & \textbf{Significant slopes} \\
& \textbf{($p < 0.05$)} & \textbf{($p < 0.05$)} \\
\midrule
Opioid dispensing & 1 & 3 \\
Adult smoking & 0 & 4 \\
Some college education & 2 & 3 \\
Unemployment & 3 & 4 \\
Primary care physicians & 0 & 1 \\
\bottomrule
\end{tabular}
\end{table}

\begin{figure}[h!]
\centering
\includegraphics[width=1\linewidth]{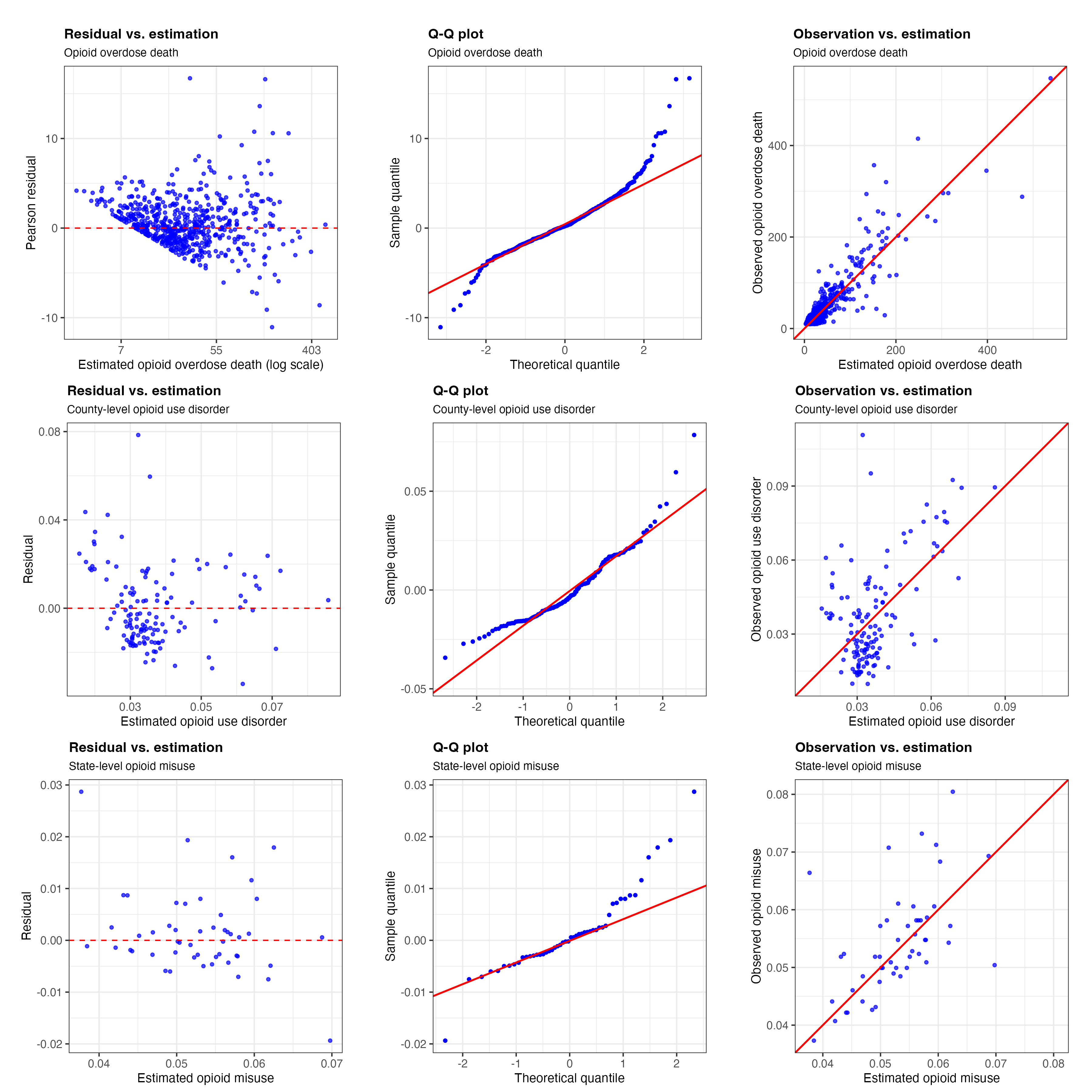}
\caption{Plots of model diagnostics for assessing the distributional assumptions and goodness-of-fit of our proposed Bayesian hierarchical model. Rows are for different input data, namely, (1) county-level overdose deaths (for counties with unsuppressed values), (2) county-level opioid use disorder prevalence estimates (in Massachusetts and Kentucky), and (3) state-level opioid misuse prevalence estimates. The columns, from left to right, show (1) residuals versus estimated values, (2) Normal Q-Q plot for the residuals, and (3) observed vs. estimated values.}
\label{fig:distributional-assumptions}
\end{figure}

\begin{figure}[h!]
\centering
\includegraphics[width=1\linewidth]{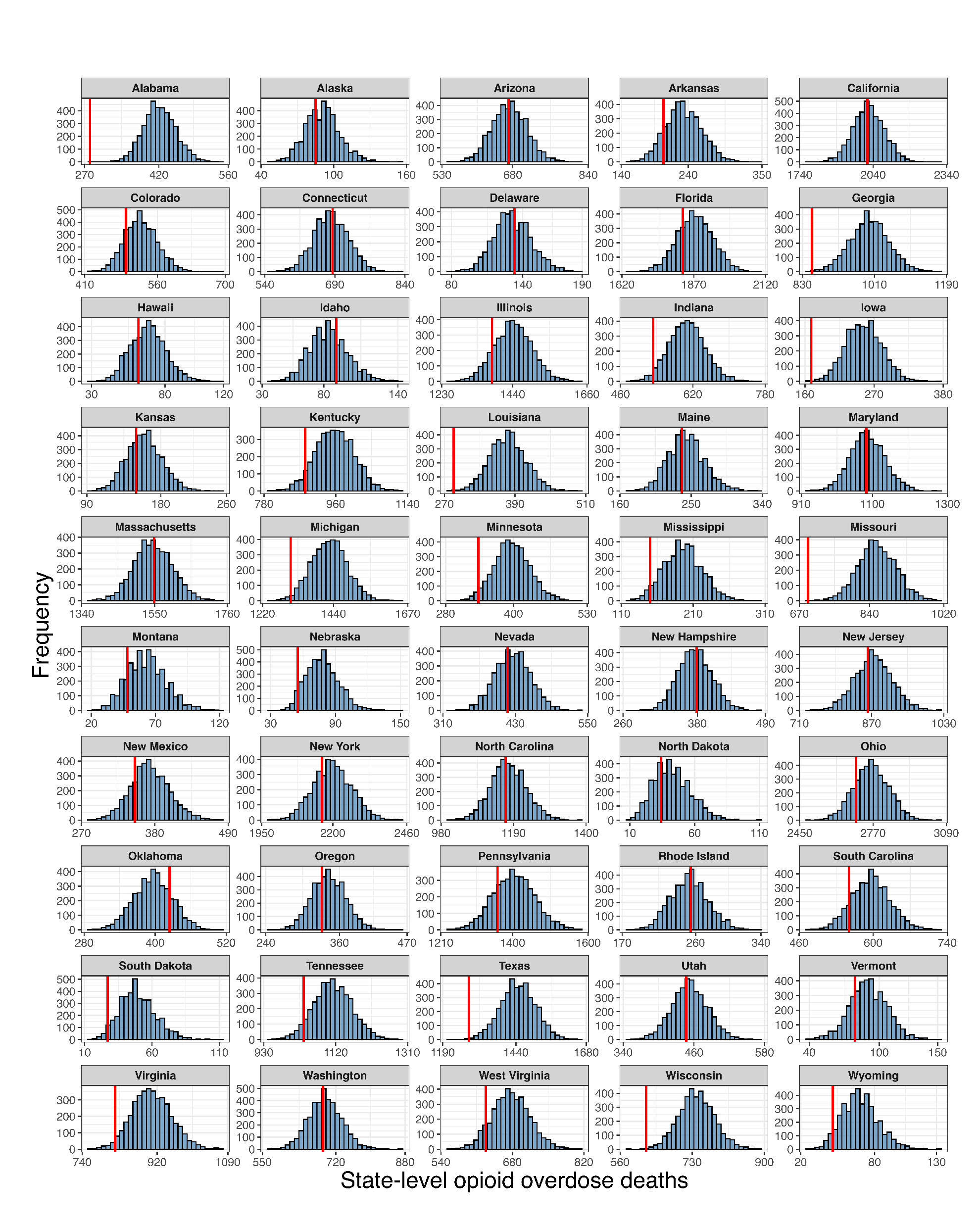}
\caption{Posterior predictive check of state-level opioid overdose deaths. The histograms represent the posterior predictive distributions of overdose deaths from the Bayesian hierarchical model by state, and the red lines represent the observed total number of overdose deaths in each state.}
\label{fig:ppc}
\end{figure}

\begin{figure}[h!]
\centering
\includegraphics[width=\linewidth]{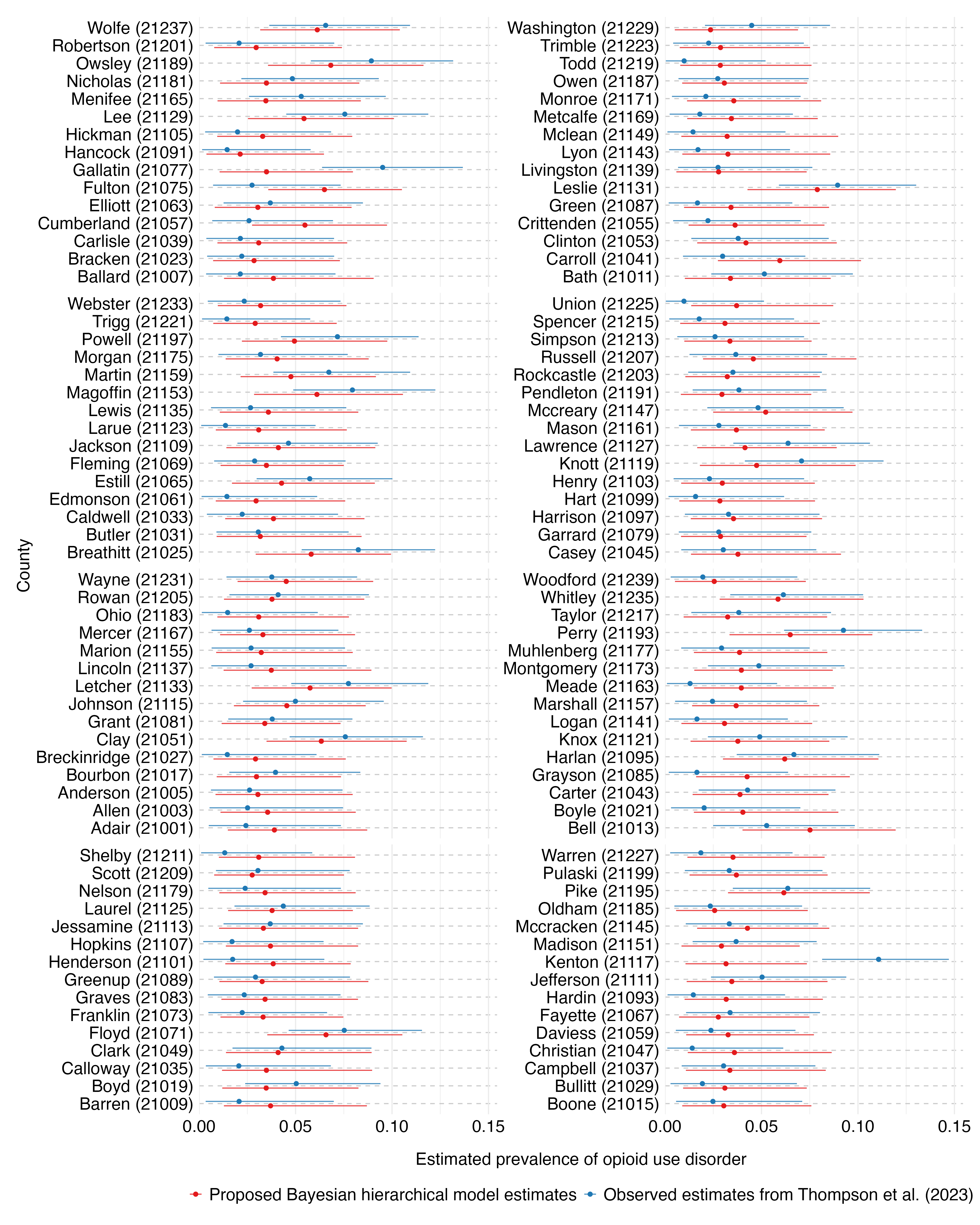}
\caption{The 10-fold cross-validation results of the Bayesian hierarchical model compared with existing estimates from \citet{Thompson2023}. The plotted credible intervals of the OUD estimates from \citet{Thompson2023} are based on the estimated $\hat{\sigma}_{KY}$ (see Section \ref{sec:real-world data settings} for more details).}
\label{fig:10foldcv-ky}
\end{figure}

\begin{figure}[t]
  \centering    \includegraphics[width=1\textwidth]{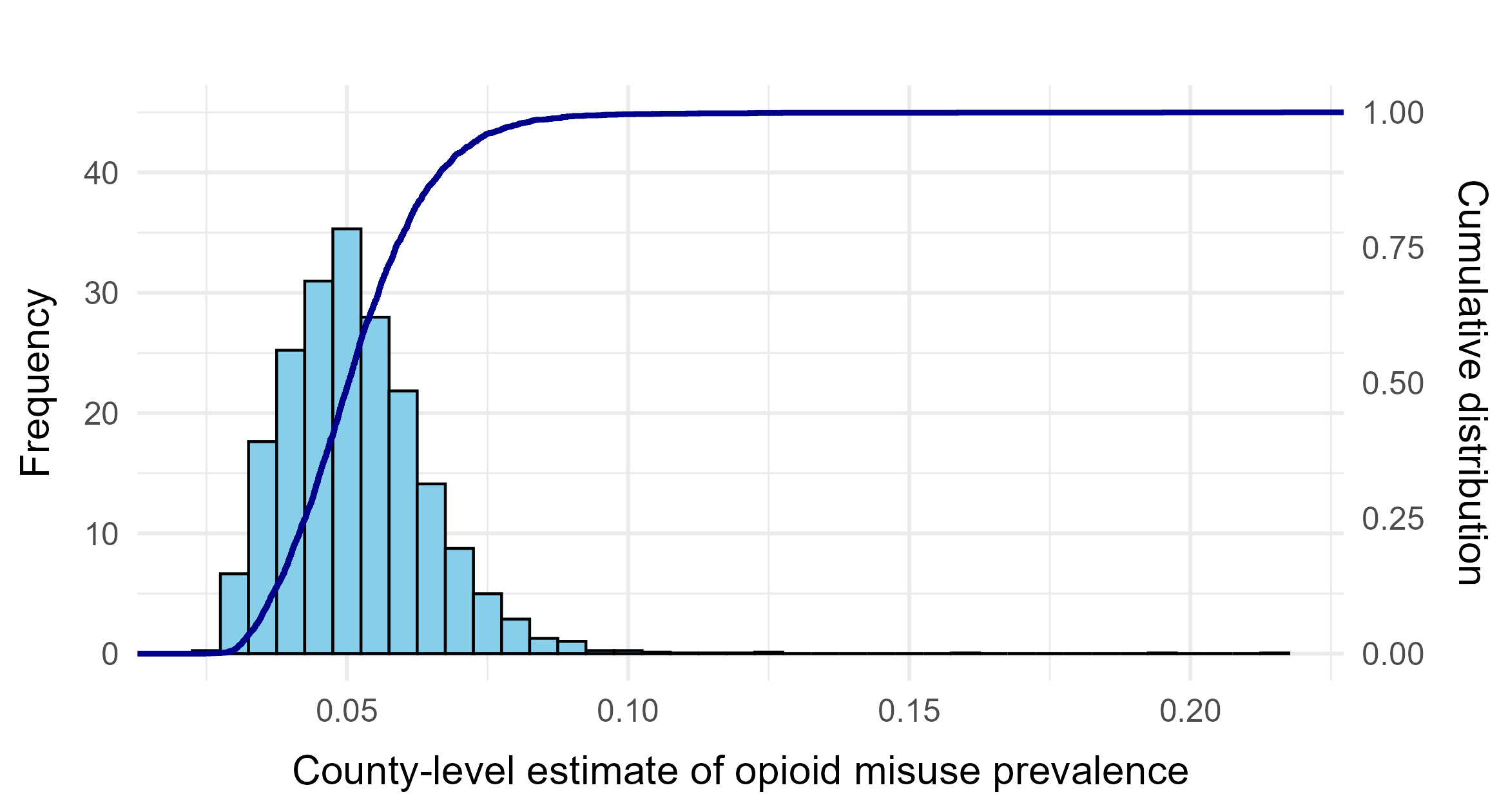}
  \caption{Distribution of estimated prevalence of opioid misuse at the county level.}
    \label{fig:histogram}
\end{figure}    

\begin{figure}[ht!]
  \centering \includegraphics[width=0.85\textwidth]{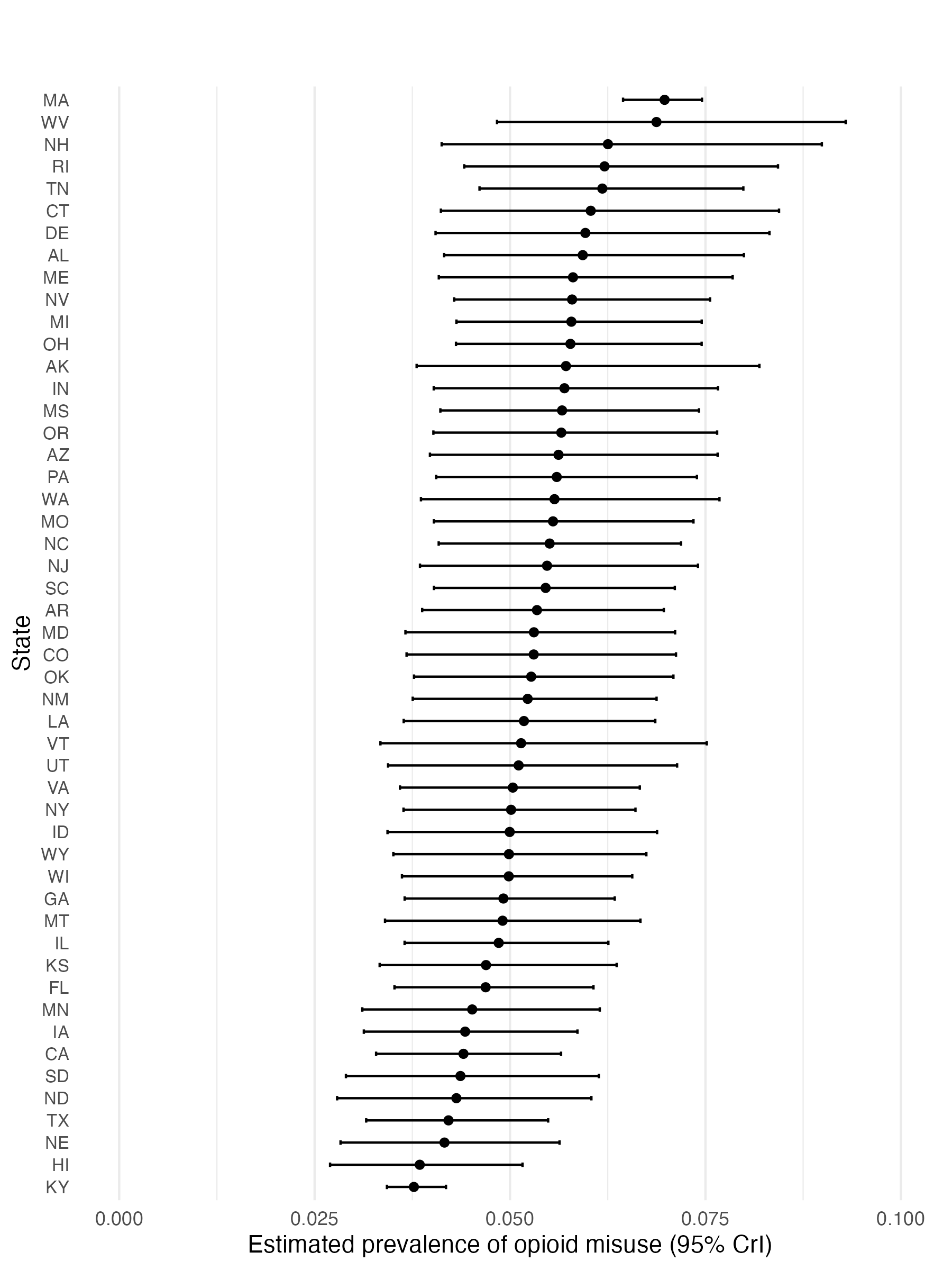}
  \caption{State-level estimates of opioid misuse prevalence.}
    \label{fig:prevalence-est-state-level}
\end{figure}

\begin{figure}[ht!]
\centering
\includegraphics[width=1\linewidth]{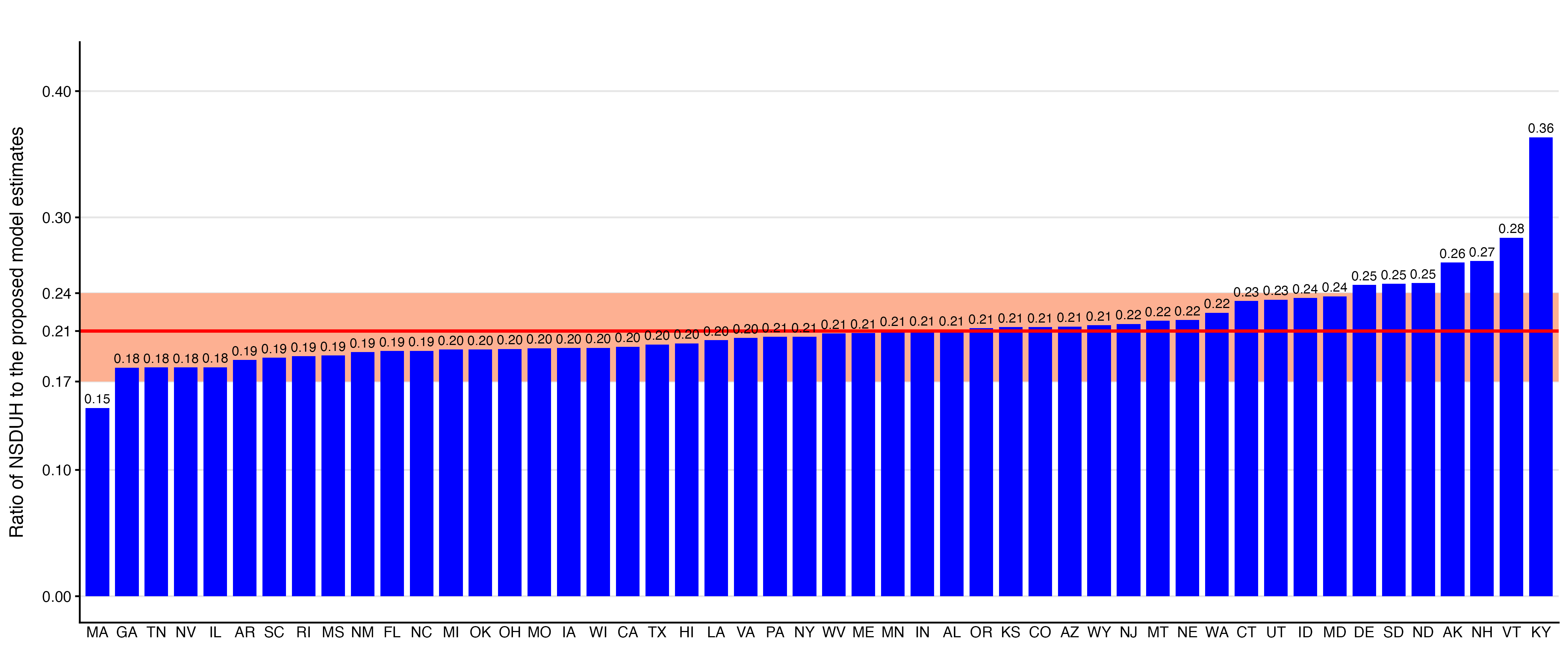}
\caption{The ratio of NSDUH state-level estimates of opioid misuse to the estimates by the proposed Bayesian hierarchical model aggregated at the state level. The red line and its shaded region indicate the posterior mean and the 95\% credible interval of the scaling coefficient $\gamma =$ 0.21 (0.17, 0.24).}
\label{fig:cmp-nsduh-state-level}
\end{figure}
\clearpage

\begin{figure}[ht!]
\centering
\includegraphics[width=\linewidth]{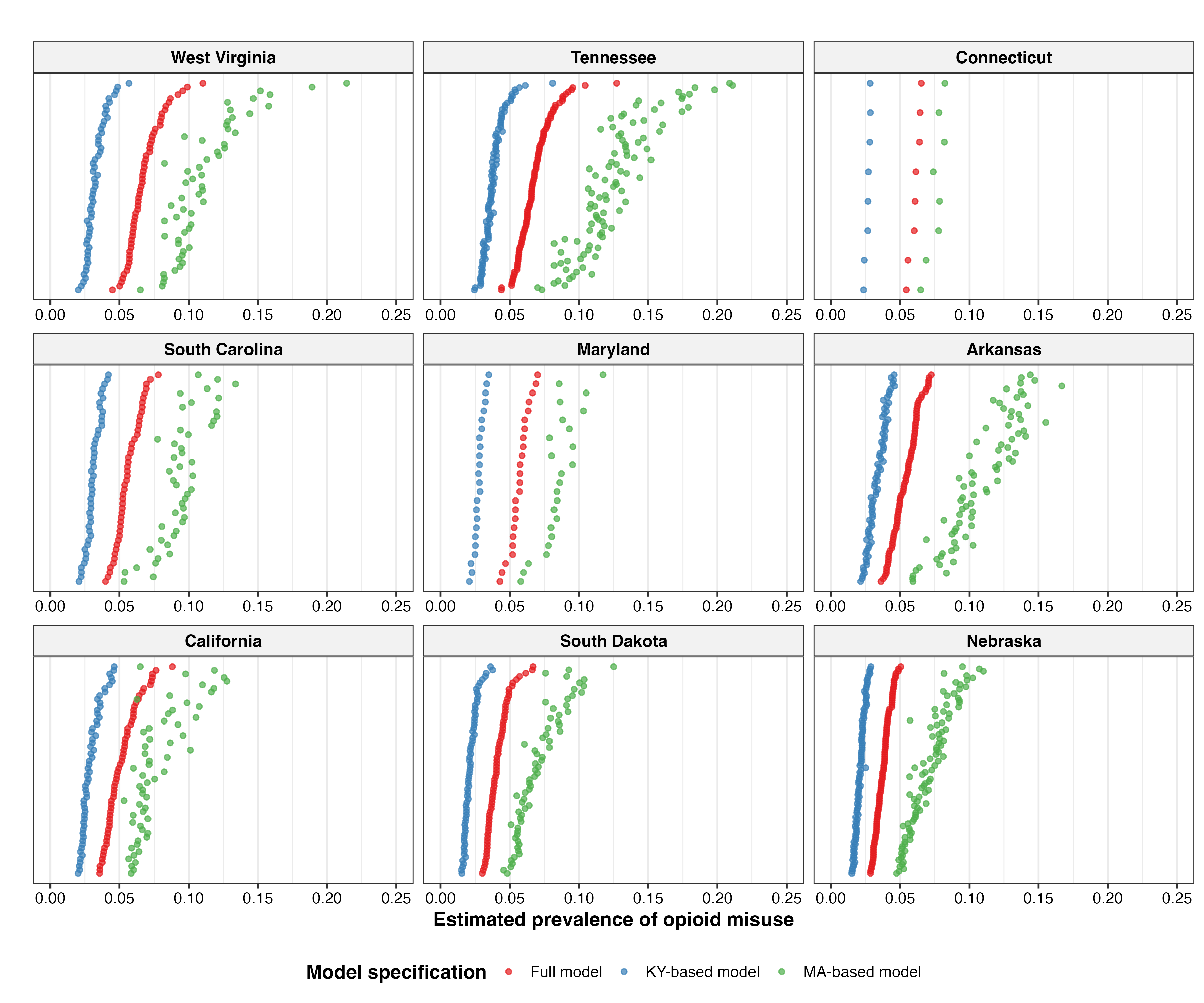}
\caption{Comparison of county-level prevalence estimates by single state-based model and the full model in the leave-one-state-out cross-validation for nine selected states. Estimates of MA- and KY-based models are obtained by holding out county-level prevalence data of Kentucky (KY) and Massachusetts (MA), respectively. For illustration purposes, nine states are selected to represent those with high (West Virginia, Tennessee, Connecticut), medium (South Carolina, Maryland, Arkansas), and low prevalence (California, South Dakota, and Nebraska). Three data points of different colors in each row represent the prevalence estimates from different models for the same county; county names are omitted in the figure for readability.}
\label{fig:loso_9states}
\end{figure}

\begin{figure} [ht!]
\centering
\includegraphics[width=1\linewidth]{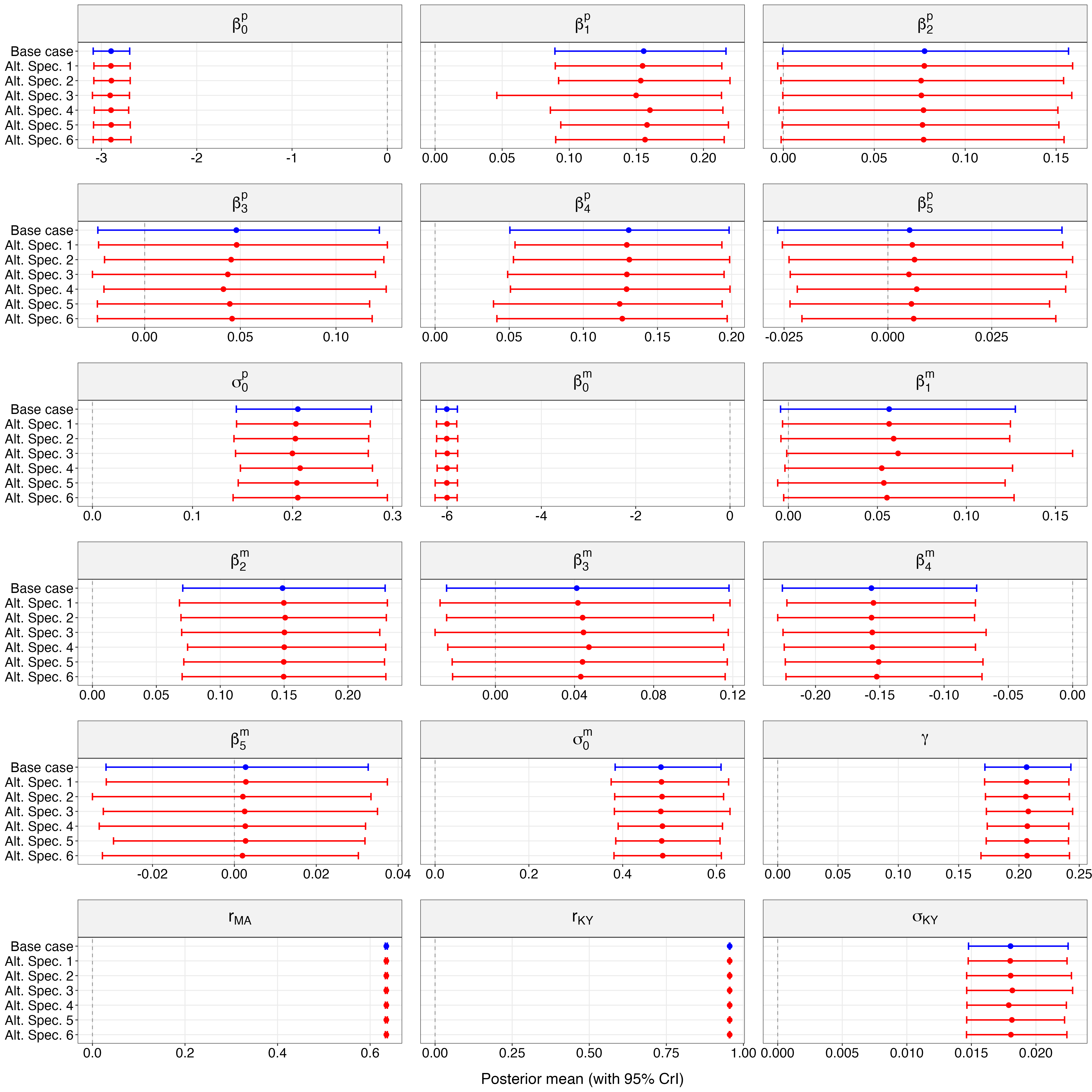}
\caption{Results of sensitivity analysis in the prior of the Bayesian model. The points and horizontal lines represent the posterior means and the 95\% credible intervals, respectively, for each parameter under each prior specification. Alternative specifications (Alt. Spec.) 1 and 2 change the prior for fixed effect intercepts $\beta_0^p$ and $\beta_0^m$ to $\mathcal{N}(0,5^2)$ and $\mathcal{N}(0,2.5^2)$, respectively (from the base case $\mathcal{N}(0,5^2)$); Alt. Spec. 3 and 4 change the prior for $\sigma_0^p, \sigma_0^m$ to Half-Normal($\sigma=5$) and Half-Normal($\sigma=2.5$), respectively (from the base case Half-Normal($\sigma=10$)); Alt. Spec. 5 applies a stronger shrinkage Half-Normal Horseshoe+ prior with scale of 0.5 (compared with 1.0 in the base case); and Alt. Spec. 6 combines Alt. Spec. 2, 4, and 5.}
\label{fig:prior_sens}
\end{figure}

\end{document}